\documentclass[10pt,twocolumn,hidelinks]{scrartcl}

\usepackage[T1]{fontenc}
\usepackage[utf8]{inputenc}
\usepackage{newtxtext,newtxmath}

\usepackage[english]{babel}
\usepackage[figurename=Fig.,labelsep=space,labelfont=bf]{caption} \setcapindent{0pt}
\usepackage{bm,dcolumn,csquotes,changepage}
\RedeclareSectionCommand[afterskip=.1\baselineskip,font=\normalfont\large\bfseries]{section}
\RedeclareSectionCommand[afterskip=.1\baselineskip,font=\normalfont\bfseries]{subsection}

\usepackage[backend=bibtex,style=phys,sortcites,biblabel=brackets,eprint=true,maxnames=5,defernumbers=true,refsegment=section]{biblatex}
\AtEveryBibitem{\clearlist{language}}

\defbibfilter{onlymain}{ segment=0 }
\defbibfilter{onlymethods}{ not segment=0 and segment=1}
\defbibfilter{onlysuppinfo}{ not segment=0 and not segment=1 }

\usepackage{mathrsfs,chemformula,graphicx,siunitx,xstring,hyperref,scalerel}
\DeclareMathOperator*{\uint}{\scalerel*{\rotatebox{8}{$\!\scriptstyle\int\!$}}{\int}}
\LetLtxMacro{\svqty}{\qty}
\usepackage{physics}
\LetLtxMacro{\qty}{\svqty}

\usepackage[left=14mm, top=17.7mm, right=12mm, bottom=18.6mm,footskip=9.3mm, a4paper]{geometry}
\newcommand{\fotonik}{Department of Photonics Engineering, DTU Fotonik, Technical University of Denmark, Building 343, DK-2800 Kgs.\ Lyngby, Denmark.}
\newcommand{\nanophoton}{NanoPhoton - Center for Nanophotonics, Technical University of Denmark, Ørsteds Plads 345A, DK-2800 Kgs.\ Lyngby, Denmark.}
\newcommand{\mekanik}{Department of Mechanical Engineering, Solid Mechanics, Technical University of Denmark, Nils Koppels Alle, Building 404, DK-2800 Kgs.\ Lyngby, Denmark.}
\newcommand{\nanolab}{DTU Nanolab, Technical University of Denmark, Building 347, DK-2800 Kgs.\ Lyngby, Denmark.}
\newcommand{\graphene}{Center for Nanostructured Graphene, Technical University of Denmark, Building 345C, DK-2800 Kgs.\ Lyngby, Denmark.}
\newcommand{\addressAffil}{
    e-mail: ssto@dtu.dk.
    \textsuperscript{1}\fotonik\
    \textsuperscript{2}\nanophoton\
    \textsuperscript{3}\mekanik\
    \textsuperscript{4}\nanolab\
    \textsuperscript{5}\graphene
}

\usepackage[activate={true,nocompatibility},final,tracking=true,kerning=true,spacing=true]{microtype}
    \microtypecontext{spacing=nonfrench}

\makeatletter
\renewcommand*{\fnum@figure}{{\normalfont\bfseries \figurename~\thefigure.}}
\renewcommand\@maketitle{
  \null
  \begin{adjustwidth}{5em}{5em}
    \begin{center}
        {\large\bfseries \@title \par  \vskip .5em}
        {\@author \par}
        {(Dated: \@date) \vskip .5em}
        {\bfseries Abstract\\}
        Nanotechnology enables in principle a precise mapping from design to device but relied so far on human intuition and simple optimizations. In nanophotonics, a central question is how to make devices in which the light-matter interaction strength is limited only by materials and nanofabrication. Here, we integrate measured fabrication constraints into topology optimization, aiming for the strongest possible light-matter interaction in a compact silicon membrane, demonstrating an unprecedented photonic nanocavity with a mode volume of $V\sim3\times10^{-4}\,\lambda^3$, quality factor $Q\sim1100$, and footprint $4\,\lambda^2$ for telecom photons with a $\lambda\sim\SI{1550}{nm}$ wavelength. We fabricate the cavity, which confines photons inside \SI{8}{nm} silicon bridges and use near-field optical measurements to perform the first experimental demonstration of photon confinement to a single hotspot well below the diffraction limit in dielectrics.
    \end{center}
    \vskip .5em
  \end{adjustwidth}
}
\makeatother

\begin{document}

\title{Nanometer-scale photon confinement in topology-optimized dielectric cavities}
\date{\today}
\author{
    Marcus~Albrechtsen\textsuperscript{1},
    Babak~Vosoughi~Lahijani\textsuperscript{1,2},
    Rasmus~Ellebæk~Christiansen\textsuperscript{2,3},
    Vy~Thi~Hoang~Nguyen\textsuperscript{4},
    Laura~Nevenka~Casses\textsuperscript{1,2,5},
    S{\o}ren~Engelberth~Hansen\textsuperscript{1,2},
    Nicolas~Stenger\textsuperscript{1,2,5},
    Ole~Sigmund\textsuperscript{2,3},
    Henri~Jansen\textsuperscript{4},
    Jesper~M{\o}rk\textsuperscript{1,2}, and 
    S{\o}ren~Stobbe\textsuperscript{1,2,}\footnote{\addressAffil}\hspace{1ex}
}

\maketitle

\begin{figure*}
    \centering
    \includegraphics[width=\linewidth]{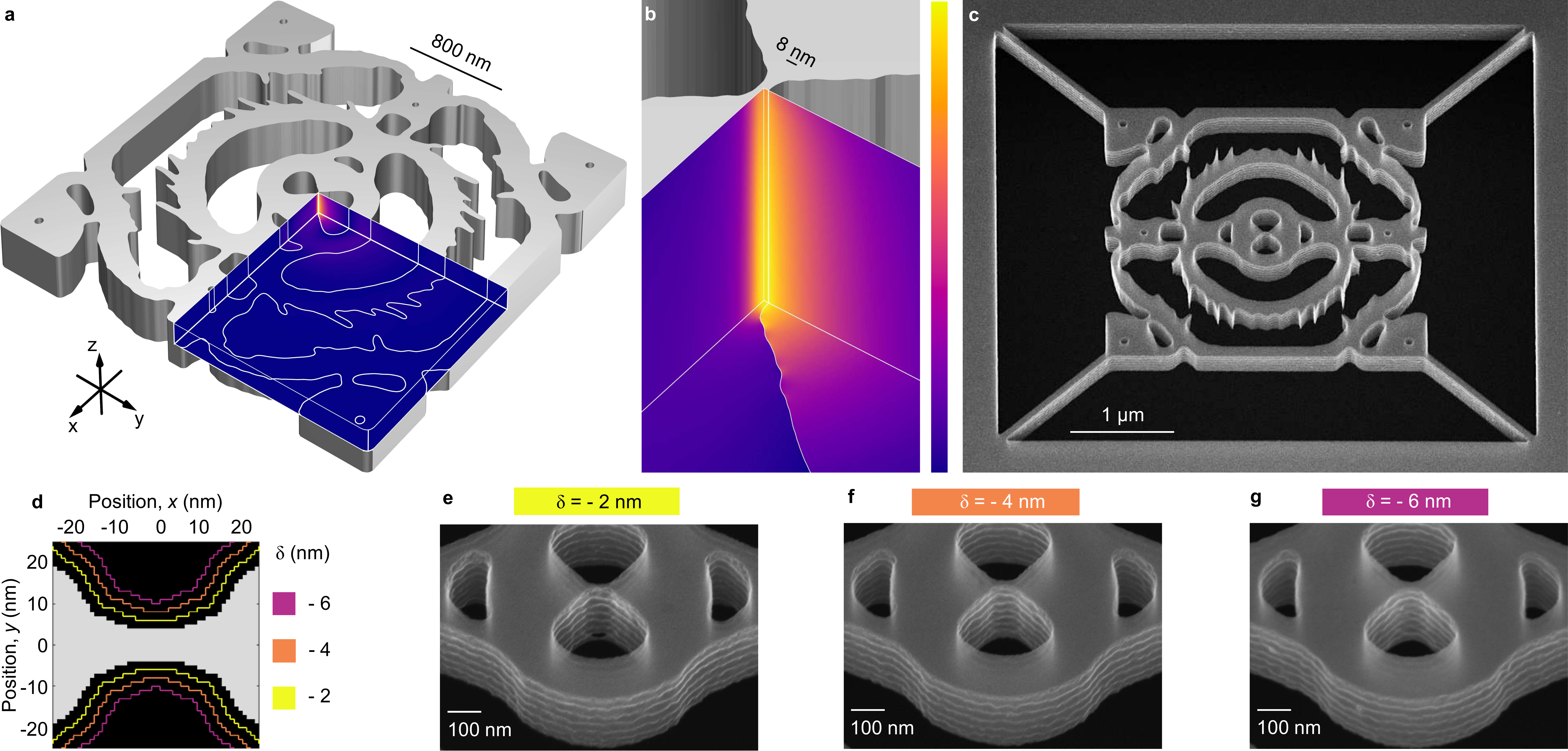}
    \caption{
    \textbf{Fabrication of topology-optimized silicon dielectric bowtie cavity (DBC).}
    (\textbf{a}) Rendering of the DBC design generated by tolerance-constrained topology optimization. The normalized $|\bm{E}|$-field is projected on the faces defining the three symmetry planes of the design.
    (\textbf{b}) Zoom-in of the solid silicon bowtie exhibiting a strong field confinement due to the bowtie bridge dimension of \SI{8}{nm}.
    (\textbf{c}) \SI{40}{\degree} tilted scanning electron microscopy (SEM) image of a fabricated cavity.
    (\textbf{d}) Global geometry-tuning, $\delta$. Each air (black) pixel ($\SI{1}{nm\squared}$) inside a $\delta$-outline is exposed uniformly with electron-beam lithography; hence, air features defining the device are uniformly tuned.
    (\textbf{e} to \textbf{g}) \SI{40}{\degree} tilted SEM images of bowtie region for $\delta=\{-2,-4,-6\}\,\SI{}{nm}$. We measure the mean width of the fabricated bowties to be $(8\pm5)\,\SI{}{nm}, (10\pm5)\,\SI{}{nm}, \text{and}\ (16\pm5)\,\SI{}{nm}$ for figures (\textbf{e}), (\textbf{f}), and (\textbf{g}), respectively, noting the variation in width along the $z$-direction caused by the scallops and $\sim\SI{1}{\degree}$ negative sidewall angle represented by the uncertainty as discussed in the main text.
    }
    \label{fig:1}
\end{figure*}

Optical nanocavities confine and store light, which is essential to increase the interaction between photons and electrons in semiconductor devices ranging from lasers to emerging quantum technologies \cite{lodahl_interfacing_2015,koenderink_nanophotonics_2015}.
A wealth of mechanisms can be exploited for building nanocavities, including
distributed Bragg reflection \cite{painter_two-dimensional_1999,akahane_high-q_2003,notomi_manipulating_2010},
total internal reflection \cite{aspelmeyer_cavity_2014},
Fano resonances or bound states in the continuum \cite{yu_demonstration_2017},
plasmonic resonances \cite{mortensen_generalized_2014}, and
topological confinement \cite{bandres_topological_2018}.
These approaches have achieved orders of magnitude improvements to the temporal confinement, but none of them allow optical mode volumes, $V$, in the deep subwavelength regime. Alternatively, plasmons in metal nanoparticles can confine light below the diffraction limit but the absorption losses in metals \cite{notomi_manipulating_2010,khurgin_how_2015} limit the quality factors to well below 100 \cite{wang_general_2006}.
The approach of our work is entirely different: We also consider loss-less dielectrics but rather than using geometry optimization of designs based on human intuition, we use geometry-agnostic inverse design, i.e., topology optimization \cite{jensen_topology_2011,aage_giga-voxel_2017,molesky_inverse_2018}, to maximize the light-matter interaction in the center of the design. Previous theoretical works \cite{gondarenko_spontaneous_2006,liang_formulation_2013,wang_maximizing_2018}
have found that this procedure results in dielectric bowtie cavities (DBCs) but inverse design is prone to yield unrealistic designs unless constrained by the limitations of materials and fabrication, which was pointed out in recent theoretical works  \cite{zhou_minimum_2015,wang_maximizing_2018}.
Here, we first measure the fabrication constraints of a state-of-the-art nanofabrication process and then include these constraints directly in the topology optimization to design a compact nanocavity. This interweaving between design and fabrication results in the novel and, importantly, realistic cavity shown in Fig.~\ref{fig:1}a and b, which we fabricate with high fidelity (Fig.~\ref{fig:1}c).
While nanotechnology based on lithography is particularly suited to accurately map designs onto fabricated devices, realizing fabrication-constrained topology-optimized devices is an outstanding challenge in inverse design that was not attempted in any field of research or engineering until now.

The underlying principle of DBCs is local field enhancements due to the electromagnetic boundary conditions across material interfaces \cite{almeida_guiding_2004,robinson_ultrasmall_2005,gondarenko_spontaneous_2006,liang_formulation_2013,wang_maximizing_2018,schneider_strong_2016,hu_design_2016,choi_self-similar_2017,albrechtsen_two_2021}. They demand that the tangential component of the electric field, $\textbf{E}$, and the normal component of the displacement field, $\textbf{D}=\epsilon\textbf{E}$, are continuous. This implies that a semiconductor bridge surrounded by void features, cf.\ Figs.~\ref{fig:1}a and b, can confine light inside the material, which is crucial to enhance the interaction with embedded emitters \cite{lodahl_interfacing_2015} or material nonlinearities \cite{choi_self-similar_2017}.
Besides the fundamentally different confinement mechanism, DBCs differ from previous cavity paradigms in several ways.
First, the small mode volume of nanometer-scale DBCs implies strong light-matter interaction without resorting to extremely high quality factors, $Q$, thus enabling applications requiring wide bandwidths such as nanoscale light-emitting diodes, few-photon nonlinearities with short pulses \cite{notomi_manipulating_2010,choi_self-similar_2017,yu_demonstration_2017}, quantum optics with broadband emitters \cite{kaupp_scaling_2013}, and optical interconnects \cite{mork_squeezing_2020}.
Second, the field enhancement of DBCs relies on the proximity to material boundaries, which implies that their modes are very sensitive to the precise size and shape of the bowtie \cite{hu_design_2016,choi_self-similar_2017,wang_maximizing_2018,mignuzzi_nanoscale_2019,zhao_minimum_2020,albrechtsen_two_2021}. Smaller bridges reduce $V$, immediately implying that a new frontier of nanocavity research is concerned with reducing the smallest feature size allowed by the nanofabrication process. This is in contrast to previous work that aimed to increase $Q$, since $V$ was believed bounded at the diffraction limit in dielectrics \cite{coccioli_smallest_1998,khurgin_how_2015}, which in turn required reducing structural disorder rather than the critical dimension \cite{minkov_automated_2014,sekoguchi_photonic_2014}.
Third, the presence of material discontinuities in few-nanometer devices makes the numerical modelling of bowtie cavities very challenging, requiring a very small mesh size \cite{choi_self-similar_2017}. Even the smallest discontinuity in the outline of the geometry implies a mode volume of zero \cite{albrechtsen_two_2021}. Such numerical artefacts arise from the well-known electric-field divergences at sharp tips and corners \cite{andersen_field_1978,landau_electrodynamics_1984,notomi_manipulating_2010,mortensen_generalized_2014,choi_self-similar_2017}. The commonly used definition of the mode volume is not generally applicable to DBCs because it can pick up such unintended surface effects rather than the effect of confining light inside the material \cite{albrechtsen_two_2021}.

Preliminary experiments in the nascent field of DBCs have so far considered bowties combined with a V-groove \cite{hu_experimental_2018} whose tip enhances the optical intensity by a lightning-rod surface effect that falls off exponentially inside the dielectric and does not provide subdiffraction confinement at the center of the cavity \cite{albrechtsen_two_2021}. The lightning-rod confinement at surfaces is not useful for semiconductor devices relying on an increased light-matter interaction inside the material and does not signify a globally confined mode. Indeed, the measurements in previous work \cite{hu_experimental_2018} found modes extending well beyond the diffraction limit. However, the V-grooves were so far inevitable because no combination of methods for lithography and etching allowed fabricating a few-nanometer high-aspect-ratio silicon bridge without eroding V-grooves into the silicon due to depletion of the etch mask and other undesirable fabrication effects.

\subsection*{Inverse design and nanofabrication}
We use carefully measured fabrication constraints as input to size- and tolerance-constrained topology optimization \cite{zhou_minimum_2015,wang_maximizing_2018} aiming to maximize the projected local density of optical states (LDOS) \cite{lodahl_interfacing_2015} at the geometric center of the domain, which is forced to be solid. The procedure for measuring the fabrication constraints is detailed in Supplementary Section~2. This ensures that the optimization is protected from local lightning-rod effects at the surface and instead achieves robust confinement inside the dielectric bowtie \cite{albrechtsen_two_2021}. Our devices are based on \SI{240}{nm} crystalline (100) silicon membranes ($n=3.48$) suspended in air, patterned with electron-beam lithography, dry etching, and selective vapour-phase hydrofluoric acid etching. We optimize a cyclic dry-etching process \cite{nguyen_core_2020} to minimize the critical dimension while tolerating periodic sidewall roughness in the form of scallops, see Methods. We note that surface roughness and the size of the scallops could be reduced by hard etching masks.
The fabrication constraints are quantified as a set of critical dimensions, which we define through minimum attainable radii. For our process, we find the radius of curvature of any solid feature, $r_s\ge\SI{10}{nm}$, and any void feature, $r_v\ge\SI{22}{nm}$.
The critical radii are limited by proximity effects during electron-beam lithography but it is possible to go below these limits with manual shape modifications of the exposure mask, see Supplementary Section~3. From systematic tests we find that it is possible to obtain a mean bowtie bridge width of \SI{8}{nm} in a localized area, which we include as a third critical radius of curvature, $r_c\ge\SI{4}{nm}$, at the center of the design domain. The topology optimization targets a maximum LDOS around $\lambda=\SI{1550}{nm}$ by tailoring the material layout in a small square domain with $2\lambda$ side length. We fabricate DBCs based on these parameters and the resulting structures show excellent agreement with the designed geometry as displayed in Fig.~\ref{fig:1}c. The high fidelity of the fabricated structures compared to the design demonstrates explicitly the value of directly including the measured fabrication constraints in the topology optimization.

The quasi-normal mode of the structure (including the tethers used to suspend the cavity, cf.\ Fig.~\ref{fig:1}c) is calculated using a finite-element method and we project the electric field $\abs{\mathbf{E}}$ on the symmetry planes of the structure in Figs.~\ref{fig:1}a and b.
We calculate the effective mode volume \cite{kristensen_generalized_2012}
\begin{equation}
    \label{eq:QNM-Vm}
    \frac{1}{V} = \Re\!\left[\frac{
    \epsilon_r(\mathbf{r}_0)\mathbf{E}(\mathbf{r}_0) \cdot \mathbf{E}(\mathbf{r}_0)
    }{
    {\displaystyle\uint\limits_V} \epsilon_r(\mathbf{r})\mathbf{E}(\mathbf{r}) \cdot \mathbf{E}(\mathbf{r})\, dV + i \frac{c\sqrt{\epsilon_r}}{2\omega} {\displaystyle\uint\limits_S} \mathbf{E}(\mathbf{r}) \cdot \mathbf{E}(\mathbf{r})\, dA
    } \!\right]\! ,
\end{equation}
with $\textbf{E}(\mathbf{r})$ and $\epsilon_r(\mathbf{r})$ the electric field and dielectric constant at position $\mathbf{r}$, respectively. $\omega$ is the complex angular eigenfrequency of the cavity mode and $c$ is the speed of light. The mode volume is in general a function of position, but for this to be a robust and useful definition, we evaluate it at the center of the cavity, $\mathbf{r}_0$. We find $V\sim0.08\ (\lambda/(2n))^3$ and $Q\sim1100$, around $\lambda=\SI{1551}{nm}$. The volume integral is over the entire simulation domain, while the surface integral should be evaluated on the outer boundaries and in practical calculations only constitutes a minor correction \cite{kristensen_generalized_2012} for cavities with $Q\gg 10$, such as our DBCs.

\begin{figure*}[t]
    \centering
    \includegraphics[width=\linewidth]{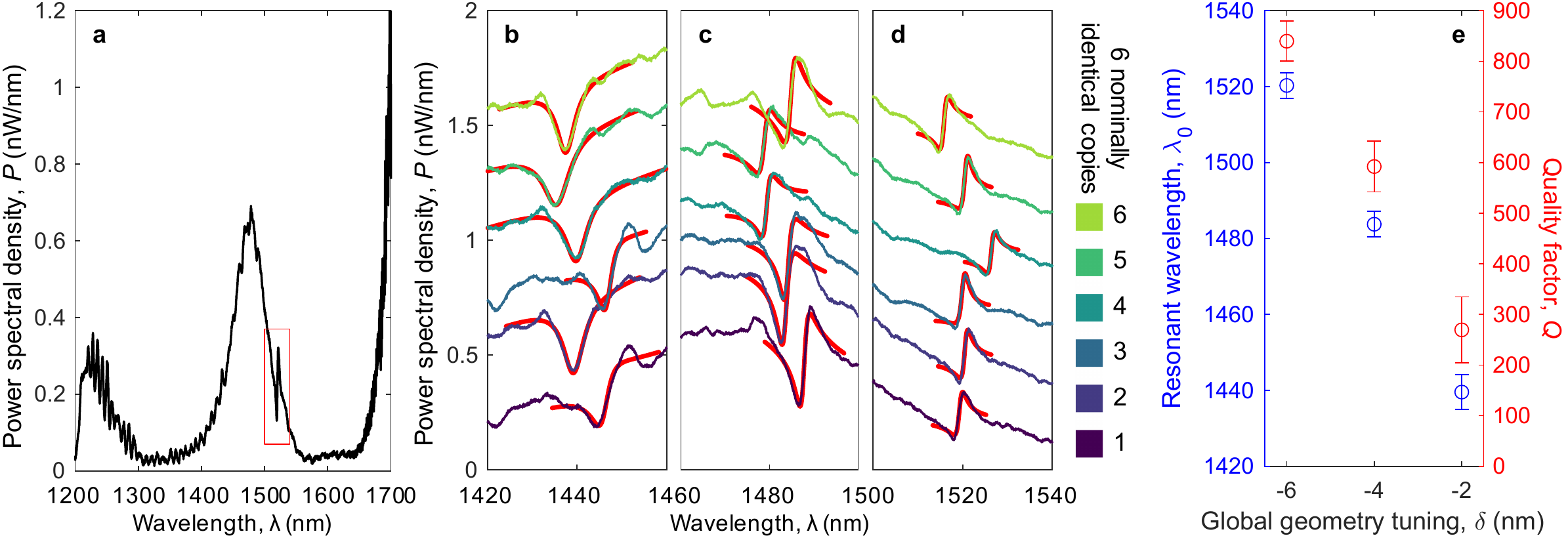}
    \caption{
    \textbf{Optical far-field characterization of dielectric bowtie cavities.}
    (\textbf{a}) Broadband spectrum of a cavity with $\delta=\SI{-6}{nm}$. The cavity mode interferes with a background mode yielding a Fano resonance centered at $\lambda\sim\SI{1520}{nm}$, highlighted by the red box.
    (\textbf{b} to \textbf{d}) Spectra for six nominally identical devices for each tuning $\delta=\{-2, -4, -6\}\,\SI{}{nm}$. The spectrum of each copy is offset incrementally by \SI{0.25}{\nano\watt\per\nm} for clarity. The full spectrum for cavity 5 is shown in (\textbf{a}). The red lines show fits to the Fano lineshape. 
    (\textbf{e}) Mean and standard deviation of resonant wavelength $\lambda_0$ (blue, left) and quality factor $Q$  (red, right) against $\delta$, extracted from the fits in (\textbf{b}-\textbf{d}).
    }
    \label{fig:2}
\end{figure*}

We stress that the bowtie, along with all other details, are emergent features arising entirely from the inverse design process. Similarly, the fact that the mode volume falls deep below the diffraction limit of $V=(\lambda/(2n))^3$ is a result of our algorithm aiming to optimize the LDOS in a limited domain. Gondarenko et al. \cite{gondarenko_spontaneous_2006} used inverse design to obtain the first DBC with confinement in air, and concluded that the bowtie shape reduces $V$ as well as that the ring gratings increases $Q$. While these features can be identified qualitatively from our inversely designed cavities, the performance of intuition-based cavity designs is inferior to topology-optimized structures \cite{wang_maximizing_2018}. Although the very large parameter space for the inverse-design algorithm makes it impossible to ascertain if the resulting design is a global optimum, it is interesting to note that the angle of the bowties are $\sim \SI{90}{\degree}$, that the bridge width equals $2r_c$, and that the voids surrounding the bridge are rounded with $\sim r_v$. These are exactly the parameters that were recently established as the global optimum for confinement of light inside bowties \cite{albrechtsen_two_2021} and the minor deviations reflect the fact that our algorithm optimizes LDOS, i.e., it targets not only the smallest $V$ but, at the same time, the largest $Q$ for the given footprint and our fabrication constraints.

Although we unambiguously demonstrate photon confinement deep below the diffraction limit, the modes are so compact that we cannot measure the precise size of the mode \cite{garcia_de_abajo_optical_2010}. Therefore, measuring the width of the fabricated silicon bridge is crucial for rigorously comparing theory and experiment for DBCs. However, the bridge width of a few nanometers is close to the resolution limit of conventional microscopy methods, such as scanning electron microscopy (SEM). We therefore fabricate three sets of DBCs, each of which subject to a global geometry-tuning, $\delta$, of the entire mask, thereby shrinking the exposed areas (air) in incremental steps of \SI{2}{nm} as shown in Fig.~\ref{fig:1}d. In order to further validate the yield and reproducibility, we fabricate and characterize six nominally identical copies of each geometry-tuned device.
Representative SEM images of each of the three geometry-tuned devices are shown in Figs.~\ref{fig:1}e to g and the \SI{2}{nm} systematic variations are clearly observed in the change of the fabricated bowtie dimensions. We measure a mean bowtie bridge width of \SI{8}{nm}, \SI{10}{nm}, and \SI{16}{nm}, for the three geometry-tuned devices, respectively. See Methods and Supplementary Section~4 for further details on the SEM characterization, and Supplementary Section~6 for an overview of devices characterized in this work.

\subsection*{Far- and near-field measurements}
We characterize the devices using confocal cross-polarized microscopy (see Methods) and a representative reflection spectrum is shown in Fig.~\ref{fig:2}a. This spectrum shows the cavity mode as a feature around \SI{1520}{nm}. The DBC mode interferes with the low-$Q$ vertical cavity mode formed by the ($\sim\SI{3}{\micro\meter}$) air gap between the silicon device layer and the silicon substrate. This results in a Fano resonance, which is well known from confocal characterization of nanocavities \cite{galli_light_2009}. The Fano line shape takes the form
\begin{equation}\label{eq:fano}
    F(\omega) = A_0(\omega) + F_0\, \frac
    {\left[ q+2(\omega-\omega_0)/\Gamma \right]^2}
    {1 + \left[ 2(\omega-\omega_0)/\Gamma \right]^2} \ ,
\end{equation}
where $\omega$ is the frequency, $\omega_0$ is the DBC resonant frequency, $\Gamma$ is the linewidth, $A_0(\omega)$ is a linear function representing the background low-$Q$ mode, $q$ measures the relative amplitudes between the main and the background modes, and $F_0$ is a constant scaling-factor. The spectra for all six copies of each of the three geometry-tuned devices shown in Figs.~\ref{fig:1}e to g are displayed in Figs.~\ref{fig:2}b to d.
We fit the Fano model locally around each resonance and extract $\omega_0$ and the quality factor $Q=\omega_0/\Gamma$ for all 18 devices of the three global geometry-tuning parameters.
Figure~\ref{fig:2}e shows the mean and standard deviation of the resonant wavelength,~$\lambda_0$, and~$Q$, for each~$\delta$.
We obtain a mean spectral shift $\Delta\lambda=(40.4\pm0.6)\,\SI{}{nm}$ between each incremental value of $\delta=-\SI{2}{nm}$ from a linear fit and find that the standard deviation of the resonance shift of each set of geometry-tuned devices is $\le(4\pm0.6)\,\SI{}{nm}$. That is, the six nominally identical copies has spectral shifts $\le\Delta\lambda/10$, which corresponds to the devices being identical within $\abs{\delta}\le\SI{0.2}{nm}$.

\begin{figure*}[!ht]
    \centering
    \includegraphics[width=\linewidth]{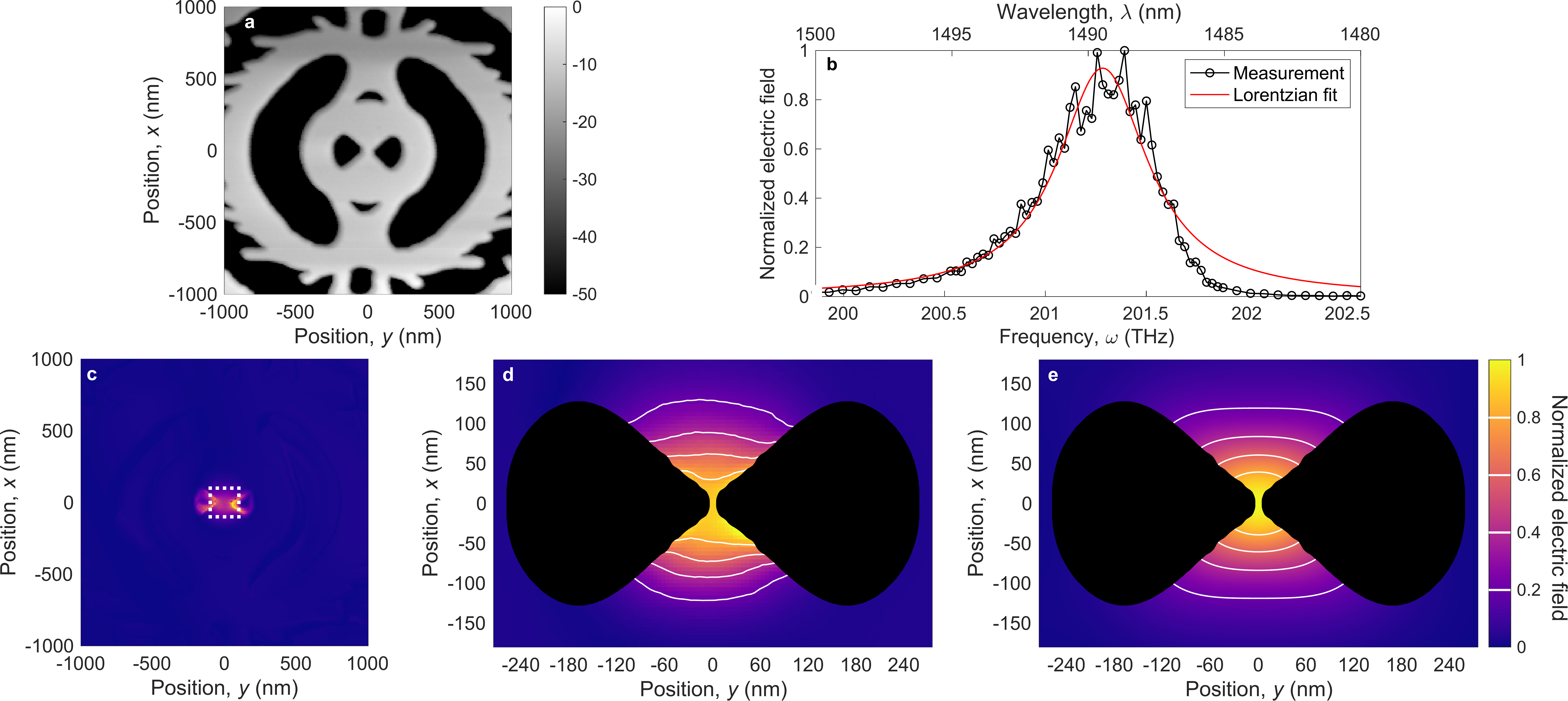}
    \caption{
    \textbf{Scattering-type scanning near-field optical microscopy (s-SNOM) of cavity mode.}
    (\textbf{a}) Topography measured by atomic force microscopy (AFM).
    (\textbf{b}) Spectrum measured of amplitude of scattered field with Lorentzian fit, $\lambda_0=(1489.4\pm 0.1)\,\SI{}{nm}$ and $Q=370\pm40$.
    (\textbf{c}) s-SNOM signal on resonance demonstrating strong field localization with excellent suppression of background noise. The white dotted box highlights a square domain with its side length given by the diffraction limit, $\lambda_0/2/n_\text{Si}\sim\SI{200}{nm}$.
    The strong signal in the cavity voids arises due to complex interactions between the AFM probe and the cavity mode.
    (\textbf{d}) Normalized measured amplitude of the light scattered from the cavity surface with the voids blacked out.
    (\textbf{e}) Numerical simulation of experiment, $f(\sigma=\SI{37}{nm}) \ast |E_c|^2$, confirming photon localization below the instrument response function as explained in the main text. The s-SNOM measurements presented here were performed on cavity copy~3 ($\delta=\SI{-4}{nm}$), which has a mean bowtie width of \SI{10}{nm}.
    }
    \label{fig:3}
\end{figure*}

While far-field measurements give important insights into the spectral properties of DBCs, they do not allow extracting information about the mode shape and confinement. We therefore interrogate the near-field immediately above the DBCs using a scattering-type scanning near-field optical microscopy (s-SNOM), where a continuous-wave laser is focused on an oscillating atomic force microscope (AFM) silicon tip scanning across the DBC.
Figure \ref{fig:3}A shows the measured topography, which provides a clear image of the device but also shows that the tip penetrates into the void features, implying that the measured geometry is convolved with the function describing the tip.
For the near-field optical characterization we use a pseudo-heterodyne interferometric detection scheme, which strongly suppresses interference with the far-field background \cite{ocelic_pseudoheterodyne_2006}.
This experiment allows recording the optical spectrum of the cavity mode without exciting the low-$Q$ background resonance. Figure~\ref{fig:3}b shows the measured amplitude at an effective height of \SI{5}{nm} above the surface at the center of the DBC.
We model the measured cavity mode using a Green-tensor formalism treating the tip as a polarizable sphere and find that the measured amplitude is modulated by the intensity of the cavity mode \cite{martin_electromagnetic_1998}.
From a Lorentzian fit in the frequency domain we obtain $\lambda_0=(1489.4\pm 0.1)\,\SI{}{nm}$ and $Q=370\pm40$. The reduction in $Q$ arises since the s-SNOM tip acts as an additional loss channel for the cavity so the s-SNOM experiment measures a loaded $Q$. The deviation from a Lorentzian lineshape may be due to nonlinear interactions or coupling with the near-field tip \cite{pellegrino_non-lorentzian_2020}.

When continuously exciting the DBC with a laser tuned to the cavity resonance while scanning the position, we can map out the spatial structure of the cavity mode. The result, which is shown in Fig.~\ref{fig:3}c, shows that the mode is strongly localized at a single hotspot. The near-field measurements in Fig.~\ref{fig:3}c show enhanced fields at the edges of the void features on the sides of the silicon bridge. These scattering fields arise because the tip goes down into the holes and therefore scatters not only surface fields but a complex combination of the surface field and the field in the voids, see Supplementary Section~5. We disregard the data obtained when the tip falls into the voids in the following analysis to facilitate a direct comparison between the measured field above the device to theoretical predictions.
Figure~\ref{fig:3}d shows a high-resolution map of the measured normalized scattered field amplitude with the regions above the voids blacked out.
The measured field cannot be compared directly to the calculated quasi-normal mode shown in Figs.~\ref{fig:1}a and b because the measured amplitude probes the intensity of the cavity mode and, in addition, because of the influence of the tip.
We model the tip instrument function, $f(\sigma)$, as a Gaussian of standard deviation $\sigma$ and maximize the overlap between measured and calculated field through the Bhattacharyya coefficient, $t = \sum \sqrt{ \abs{\mathbf{E}_s}\,\cdot (\abs{\mathbf{E}_c}^2 \ast f(\sigma) ) }$,  where $\ast$ denotes convolution and $\mathbf{E}_s$ ($\mathbf{E}_c$) is the measured (calculated) field $\SI{15}{nm}$ above the surface. The tip has a nominal radius of curvature of \SI{10}{nm} and probes the field when the edge is \SI{5}{nm} above the surface. Both fields are normalized, $\sum\abs{\mathbf{E}_s}=\sum\abs{\mathbf{E}_c}^2 \ast f(\sigma) =1$, with the sum being over all pixels. This analysis yields $\sigma=(37\pm5)\,\SI{}{nm}$ and $t=0.984$ indicating an excellent agreement between theory and experiment. The instrument function has a full width at half maximum of $2\sqrt{2\log(2)}\sigma=\SI{87}{nm}$, which is broader than the DBC mode size, so the measurement gives an upper bound to the mode volume below the diffraction limit, $V=(\lambda/(2n))^3$. This corresponds to a cube with a side length of $\sim\SI{200}{nm}$ as indicated by the dotted white box in Fig.~\ref{fig:3}c. Notably, even after the expansion of the mode above the cavity and after broadening by the instrument function of the near-field tip, the raw data shows an optical mode confined well below the diffraction limit. Additional s-SNOM measurements (see Supplementary Section~5) on a device of different global geometry-tuning, $\delta=\SI{-6}{nm}$ (corresponding to a mean bowtie width of \SI{16}{nm}), also yields the largest overlap, $t=0.991$, for the same instrument function $\sigma = \SI{37}{nm}$. The overlap between the two measurements is $t=0.996$, which further confirms that the DBC mode is localized below the instrument function. These results constitute the first direct experimental measurement of subdiffraction confinement of light in a dielectric structure.

\subsection*{Conclusion}
Strongly confining light inside dielectrics, as opposed to in air, vacuum, or at material boundaries, is central to applications relying on enhancing the light-matter interaction. Our work demonstrates for the first time the advantages of including measured fabrication constraints in topology optimization. This sets a new standard for photonic nanotechnology in the quest for globally optimal structures \cite{zhao_minimum_2020,chao_physical_2021} and demonstrates for the first time photon confinement inside dielectrics below the diffraction limit without intrinsic limits on $Q$. The directly optimized LDOS of our cavity corresponds to an enhancement of the light-matter interaction by a Purcell factor \cite{notomi_manipulating_2010} of $6 \times 10^{3}$ over a bandwidth of up to \SI{2}{nm}. This large bandwidth is needed for nonlinear optics and optical interconnects \cite{mork_squeezing_2020} and appears on purpose in our design due to the compact device footprint \cite{liang_formulation_2013,wang_maximizing_2018} of $4\lambda^2$. Extending the design domain would result in much higher $Q$, and our work therefore not only demonstrates unprecedented levels of photon confinement inside dielectrics, it also paves the way for experiments in extreme regimes of light-matter interaction, which in turn can suppress quantum decoherence due to phonons \cite{denning_optical_2020}.

We note that the semiconductor technology nodes, such as the current "5-nm node", of the semiconductor industry no longer describe the smallest features in integrated circuits defined by lithography \cite{murmann_nano-chips_2020}. In fact, the current industry roadmap for lithography does not aim to go below \SI{8}{nm} before 2034. The ability to fabricate highly optimized devices with \SI{8}{nm} dimensions and high aspect ratios is therefore unlocking new experimental regimes throughout most areas of semiconductor nanotechnology \cite{rogers_synthesis_2011}, including nanophotonics \cite{koenderink_nanophotonics_2015}, cavity optomechanics \cite{aspelmeyer_cavity_2014},
nanoelectromechanics \cite{midolo_nano-opto-electro-mechanical_2018}, and
quantum photonics \cite{lodahl_interfacing_2015}.

\printbibliography[segment=\therefsegment,filter=onlymain]

\section*{Methods}

\subsection*{The inverse design process}
For the inverse design procedure we model the physics using Maxwell's equations in a finite volume of space, assuming time-harmonic field behavior. We exploit the three-fold spatial symmetry of the DBC structure to reduce the model size and truncate the modelling domain using symmetry conditions and first-order absorbing boundary conditions \cite{wang_maximizing_2018}. The model is discretized and solved using the finite-element method with first-order Nedelec elements \cite{jin_finite_2015}.
The problem of designing a DBC is solved using topology optimization by recasting it as a continuous constrained optimization problem \cite{christiansen_inverse_2021}. In this process we select a subset of the model domain, i.e., the design domain, and introduce one spatially constant continuous design variable per finite element in the design domain. We apply a filtering and thresholding procedure \cite{wang_projection_2011,zhou_minimum_2015} to regularize the design. The filtered and thresholded design variables are linked to the model through a material interpolation scheme \cite{christiansen_non-linear_2019}. Hereby, the design variables control the material distribution. The optimization problem is solved using the globally convergent method of moving asymptotes \cite{svanberg_class_2002}. 
For the domain considered in this work, we choose a fixed membrane thickness of \SI{240}{nm} and restrict the design to only vary in the $(x,y)$-plane by linking the design variables along the $z$-direction \cite{christiansen_compact_2021,jensen_topology_2011}. Before the design process is executed, we specify the design domain, the measured minimum radii of curvature of the solid and void phases in the design as well as at the center, and further specify the targeted cavity-resonance wavelength and the position of the mode extremum in the cavity. Otherwise, we allow the design to emerge freely from the design process.

\subsection*{Fabrication processes}
A 25-by-\SI{25}{mm} chip is cleaved from a silicon-on-insulator wafer with a \SI{240}{nm} (100) device layer and a \SI{3}{\micro\meter} buried oxide. It is cleaned sequentially with de-ionized water, acetone, isopropanol (IPA), and dried with dry N$_2$.
The sample is dehydrated for \SI{5}{min} at \SI{200}{\celsius} and $\sim\SI{65}{nm}$ chemically semi-amplified resist (CSAR) is spin-coated from CSAR6200.04 (CSAR6200.09 diluted 1:1 in anisole) at \SI{6000}{rpm} for \SI{60}{s} followed by a \SI{5}{min} softbake at \SI{200}{\celsius}.
Six nominally identical copies of the cavity layout (56 combinations of local mask corrections and global geometry-tuning) are exposed uniformly on a \SI{100}{keV} \SI{100}{MHz} JEOL-9500FSZ electron-beam writer with current $I=\SI{202}{pA}$, dose density $D_0=\SI{3}{\atto\coulomb\per\nano\meter\squared}$, and shot pitch, $p=\SI{1}{nm}$.
The samples are developed for \SI{60}{s} in AR-600-546 (amyl acetate), cleaned in IPA, and dried with dry N$_2$ in an automatic Laurell EDC 650 puddle developer for high reproducibility.
All devices are separated by \SI{25}{\micro\meter} to reduce proximity effects.

The patterns are transferred to the device layer with 10 cycles of a modified version of the CORE-sequence \cite{nguyen_core_2020} operated at $+\SI{20}{\celsius}$. This process is a low-power switched reactive ion etching process using \ch{SF6} for the etch and oxygen for sidewall passivation, thus avoiding fluorocarbon residues.
Specifically, we fine-tune the process to achieve an aspect ratio of 30 from the thin softmask required by lithography. To improve the mask selectivity we reduce the platen power of the R-step from $\SI{10}{W}\rightarrow\SI{8}{W}$ and to reduce sidewall erosion we increase the O-step (passivation) from $\SI{3}{s}\rightarrow\SI{4}{s}$. Lastly, we reduce the \ch{SF6} flow in the E-step from $\SI{15}{sccm}\rightarrow\SI{10}{sccm}$, and modify the duration of this step from $\SI{73}{s}\rightarrow\SI{72}{s}$.
The resist is removed with 1165 Remover (N-Methyl-2-pyrrolidone) followed by IPA and dried with dry N$_2$. The sample is then cleaned for \SI{10}{min} in a Tepla 300 barrel asher with \SI{400}{sccm} \ch{O2}-flow and \SI{70}{sccm} \ch{N2}-flow at \SI{1}{kW} reaching a maximum temperature of \SI{72}{\celsius}.
The buried oxide is etched in anhydrous hydrofluoric-acid (\SI{99.995}{\percent}) vapour using ethanol as catalyst at a process pressure \SI{131}{Torr} in an SPTS Primaxx uEtch enabling both pressure and temperature control throughout the release. The sample is baked for \SI{5}{\minute} at \SI{200}{\celsius} prior to the release etch to avoid residues.

\subsection*{Scanning electron microscope characterization}
We measure the dimensions of the fabricated structures by comparing a combination of top-view and tilted SEM images analyzed with detailed image analysis, presented and discussed in Supplementary Section~4. We measure the width of the bowties as \SI{13}{nm}, \SI{15}{nm}, and \SI{21}{nm} from the top-view SEM images of the 3 sets of devices presented in Fig.~\ref{fig:1}e to g, respectively. Furthermore, we use multiple tilted views to estimate the width at the bottom of the bowties, which we find is $\sim\SI{10}{nm}$ narrower than at the top. This implies a negative sidewall angle $\sim\SI{1}{\degree}$ of all devices and a mean width of the bowtie bridges of \SI{8}{nm}, \SI{10}{nm}, and \SI{16}{nm}, for the three geometry-tuned devices, respectively, consistent with the critical radius of curvature imposed on the topology optimization.
Supplementary Section~1 presents careful numerical simulations of the fabricated dimensions, which both includes the sidewall angle as well as variations of the dimensions of the calculated structure. This confirms that the mode volume in the center of our tolerance-constrained DBC-design remains robust to variations and is deep below the diffraction limit.

\subsection*{Confocal cross-polarized microscopy setup}
A supercontinuum laser (NKT Photonics SuperK Compact) is focused on the cavity through a $\text{NA}=0.4$ microscope objective. The scattered light is collected through the same objective and measured with an optical spectrum analyzer (AQ6370D Yokogawa), wavelength range $\lambda=[1200,1700]\SI{}{nm}$. The excitation polarization is controlled with a $\lambda/2$-plate and light is collected through a linear polarizer rotated \SI{90}{\degree} to reduce specular reflections. Both excitation and collection is rotated \SI{45}{\degree} to the main optical axis of the cavity (along $x$ in Fig.~\ref{fig:1}a).

\subsection*{Near-field optical measurements}
We use an s-SNOM (Neaspec, neaSNOM), equipped with a pseudo-heterodyne module, in reflection mode to map the DBC modes in the near-field. The incident light from a tunable continuous-wave laser (Santec, TSL-710) is focused on a silicon AFM probe (NanoWorld, Arrow-NC) with a nominal tip radius of \SI{10}{nm}. The probe is used in intermittent contact mode at a frequency $f_0=\SI{280}{kHz}$ oscillating with an amplitude of \SI{60}{nm}. The amplitude of the scattered signal depends nonlinearly on the height above the sample due to the near-field contribution, therefore, demodulating at $4 f_0$ with a lock-in amplifier yields the near-field signal at the smallest height ($\sim\SI{5}{nm}$) above the surface while strongly suppressing contributions from the far-field background. The laser is s-polarized, which is aligned along the $x$-axis of the DBC (see Fig.~\ref{fig:1}a), to minimize the perturbation from the tip and to excite the cavity most efficiently. A polarizer is placed in front of a photoreceiver (New Focus, 2053-FS) to select the s-polarization of the scattered field.
We determine the resonant wavelength in the near-field from a Lorentzian fit to the near-field spectrum obtained from a fixed position in several spatial maps obtained around the bowtie for a number of wavelengths in a \SI{20}{nm} band, see Supplementary Section~5 for further details.

\printbibliography[segment=\therefsegment,filter=onlymethods]

\section*{Acknowledgments}
We gratefully acknowledge financial support from the Villum Foundation Young Investigator Program (Grant No. 13170) as well as the Experiment Program (Grant No. 00028233), the Danish National Research Foundation (Grants No. DNRF147 - NanoPhoton and No. DNRF103 - CNG), the Independent Research Fund Denmark - Natural Sciences (projects no. 0135-004038 and 0135-00315), and Innovation Fund Denmark (Grant No. 0175-00022 - NEXUS). We thank Fengwen Wang for insightful discussions about the inverse design process. We acknowledge Philip T. Kristensen for valuable discussions and input to the theoretical models.

\section*{Author contributions}
S.S. and J.M. initiated and supervised the project.
M.A., supervised by B.V.L and S.S., developed the lithography concepts, fabricated the sample, performed SEM characterization, far-field optical characterization, and most of the data analysis.
V.T.H.N. and H.J. developed the dry-etching process.
R.E.C. and O.S. developed and carried out the topology optimization.
L.N.C. and N.S. performed the near-field measurements.
M.A., B.V.L., R.E.C., S.E.H., L.N.C., N.S., O.S., J.M., and S.S. contributed to data analysis, discussions, and preparation of figures.
M.A., B.V.L., and S.S. wrote the manuscript with contributions and input from all authors.

\section*{Data availability}
Data is available upon reasonable request.

\section*{Competing financial interests}
The authors declare no competing financial interests.

\newgeometry{left=1in, top=17.7mm, right=1in, bottom=18.6mm,footskip=9.3mm}
\onecolumn

\begin{center}
    \bfseries\Large Supplementary information for \\
    \bfseries\large Nanometer-scale photon confinement in topology-optimized dielectric cavities
\end{center}

\setcounter{figure}{0}
\setcounter{equation}{0}
\makeatletter
    \renewcommand{\theequation}{S\arabic{equation}}
    \renewcommand{\thefigure}{S\arabic{figure}}
\makeatother

\section{Numerical simulation of dielectric bowtie cavities}
The mode volume is defined as the inverse of the normalized energy density evaluated at a position $\mathbf{r}_0$. This follows, e.g., from the Purcell factor describing the enhancement of the decay rate in a nanostructure compared to a homogeneous medium, i.e., the ratio of the local density of optical states (LDOS) to the density of optical states (DOS) \cite{lodahl_interfacing_2015,kristensen_generalized_2012,sanders_analysis_2018}. For nanocavities with high $Q$ and/or small $V$ it is often an excellent approximation to assume that the LDOS is dominated by a single cavity mode, thus neglecting continuum modes. The LDOS is a function of frequency, polarization, and position and these dependencies remain crucial in the single-mode approximation. The mode is a quasi-normal mode and can be calculated in several ways, for example by exciting the cavity with a dipole at position $\mathbf{r}_0$ and computing the LDOS directly \cite{liang_formulation_2013,wang_maximizing_2018}, which is the method we employ in our topology optimization.
The quasi-normal mode can also be calculated from the eigenfrequency, which is what we employ everywhere else in this work. The eigenfrequency is a linear solution, which can be scaled arbitrarily and therefore it must be normalized to yield an absolute energy density. This can be done in several ways \cite{kristensen_normalization_2015,kristensen_modeling_2020}, and in practical calculations it is convenient to use Eq.~(1) in the main text.

\begin{figure}[!b]
    \centering
    \includegraphics[width=.5\linewidth]{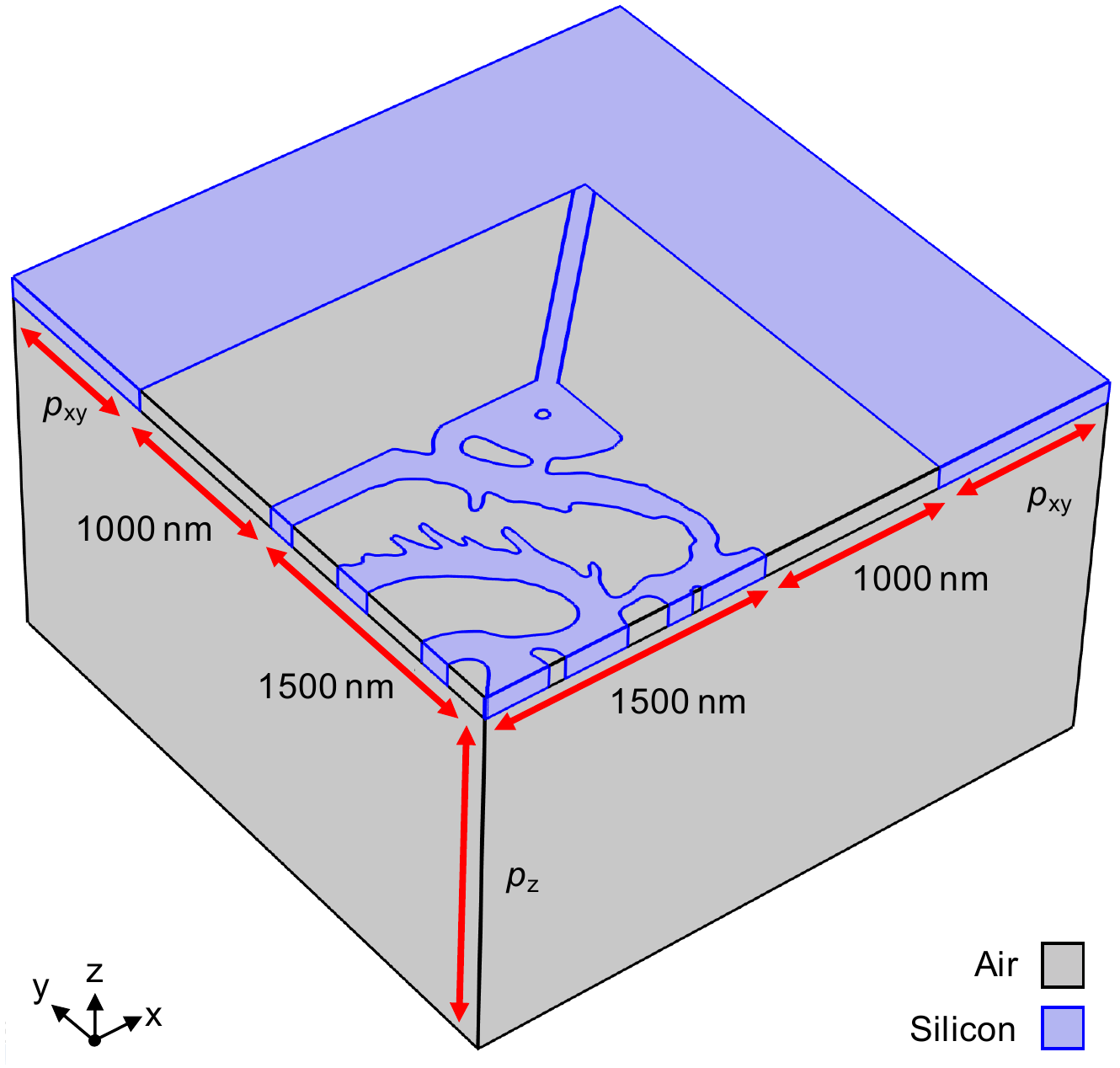}
    \caption{
    \textbf{Simulation domain for a DBC.}
    The blue domains are silicon and the gray domains are air with the in-plane padding $p_\text{xy}=\SI{1.5}{\micro\meter}$ and the out-of-plane padding $p_\text{z}=\SI{3.1}{\micro\meter}$. The $xy$-plane and $xz$-planes are perfect magnetic-conductor boundary conditions, and the $yz$-plane is a perfect electric-conductor boundary condition. For the angled-sidewall calculation, we do not employ the boundary condition on the $xy$-plane as this symmetry is broken. First-order absorbing boundary conditions are applied on the boundaries facing away from the cavity and are used for the surface integral when computing the mode volume as detailed in the main text.
    }
    \label{fig:SI-1:1}
\end{figure}

We use finite-element modelling \cite{jin_finite_2015} to simulate our structures. Figure~\ref{fig:SI-1:1} shows the numerical model solved in a finite-element model with COMSOL Multiphysics 5.6. Only 1/8th of the dielectric bowtie cavity (DBC) is simulated due to symmetries, except for our simulations of a negative sidewall angle, which breaks the out-of-plane symmetry enabling only two symmetry planes. The structure is meshed using a different mesh size around the central bowtie and the mesh size in all regions is determined by convergence tests, which results in a finer mesh around the bowtie.

\begin{figure}
    \centering
    \includegraphics[width=\linewidth]{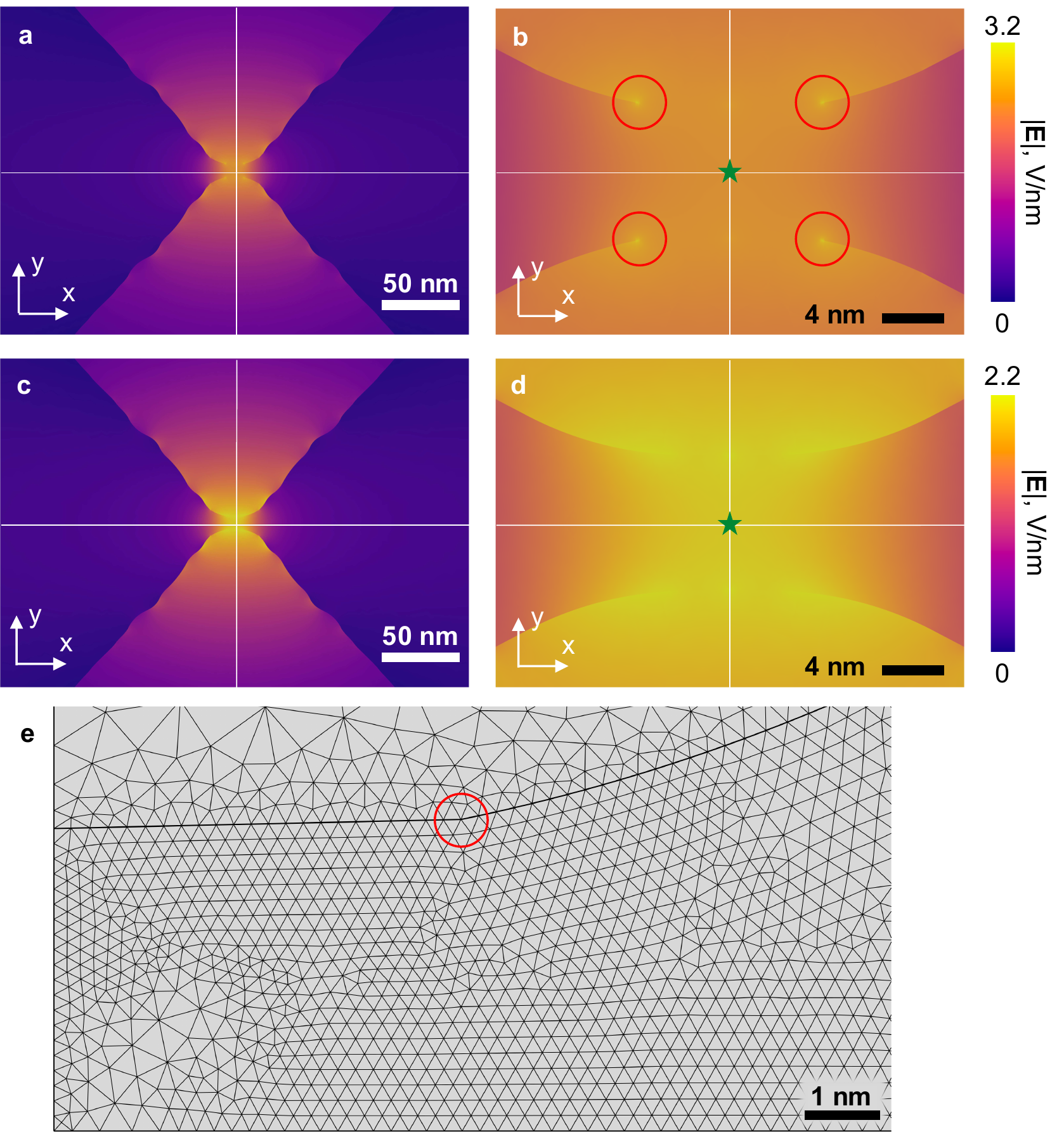}
    \caption{
    \textbf{Numerical challenges and local lightning-rod effects in DBCs.}
    \textbf{a}, Numerical evaluation of the DBC where polygon vertices are not rounded to higher-order nodes in the geometry, and \textbf{b}, zoom-in around the bowtie with red circles indicating the lightning-rod enhancements of $\abs{\mathbf{E}}$. \textbf{c}, The same geometry evaluation with rounding of the problematic vertices. \textbf{d}, This results in a continuous field whose intensity at the center of the geometry is close to the maximum at the interface. The color scales are normalized by the peak field amplitude in \textbf{a}-\textbf{d}, and the field profiles are not smoothed. The white lines shows the symmetry planes through $x=0$ and $y=0$. \textbf{e}, The mesh of 1/4 of the structure around the bowtie with the problematic mesh elements highlighted (red circle).
    }
    \label{fig:SI-1:2}
\end{figure}

\begin{figure}
    \centering
    \includegraphics[width=\linewidth]{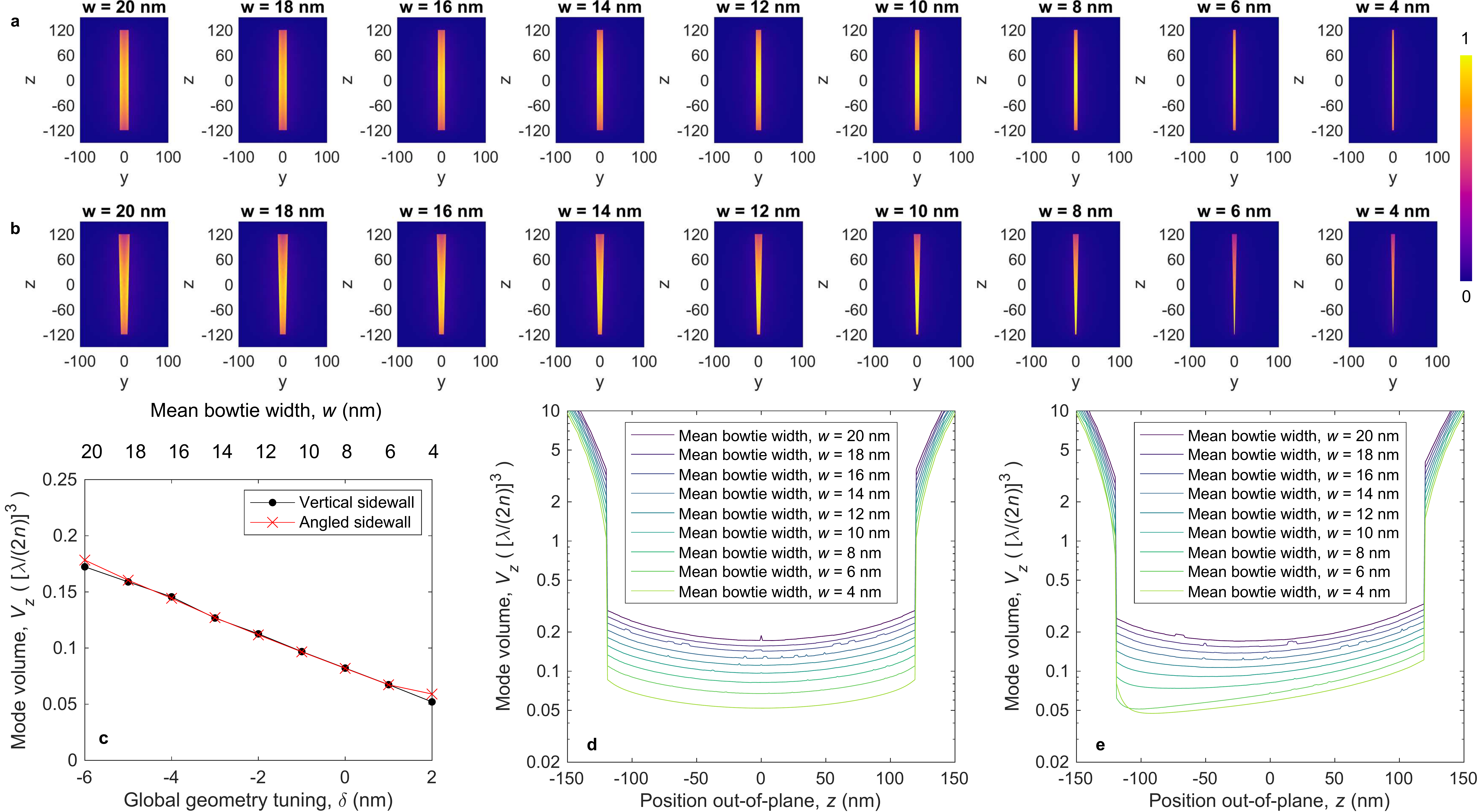}
    \caption{
    \textbf{Simulations of mode volume dependence on bowtie width and sidewall angle.}
    \textbf{a}, Calculated electric energy $\epsilon(r)\abs{\mathbf{E}(r)}^2$ in the $yz$-plane at $x=\SI{0}{nm}$ as a function of bowtie width, $w$, assuming a vertical sidewall.
    \textbf{b}, Calculated electric energy in the $yz$-plane at $x=\SI{0}{nm}$ as in \textbf{a} with a non-vertical sidewall where the width is $\SI{5}{nm}$ more narrow (wide) at the bottom (top).
    \textbf{c}, Effective mode volume according to Eq.~(1) in the main text calculated for each of the 18 devices in \textbf{a} and \textbf{b} and mapped against global geometry-tuning, $\delta$, and mean bowtie width.
    \textbf{d}-\textbf{e}, Effective mode volume evaluated for different positions along the $z$-axis for vertical (\textbf{d}), and non-vertical (\textbf{e}) sidewalls and the different bowtie widths shown in \textbf{a}-\textbf{b}. 
    }
    \label{fig:SI-1:3}
\end{figure}

Figure~\ref{fig:SI-1:2} shows the results of the numerical simulations of the DBC design. Rather than the Manhattan design outline with all vertices at \SI{90}{\degree} to control the discretization of the electron-beam lithography, a smoothed outline is extracted from the topology optimization for the numerical calculations \cite{zhou_minimum_2015,wang_maximizing_2018}.
Even if the polygons are rendered with high resolution, the discontinuities at vertices will introduce numerical artefacts as shown in Figs.~\ref{fig:SI-1:2}\textbf{a}-\textbf{b} where, e.g., the spots highlighted by red circles contains dielectric wedges, which will diverge with finer mesh \cite{sommerfeld_mathematical_2004,landau_electrodynamics_1984,jackson_classical_1999,andersen_field_1978,van_bladel_field_1985,mortensen_generalized_2014,choi_self-similar_2017} resulting in an apparent enhancement of the mode volume locally, here by $\sim\SI{50}{\percent}$. It is therefore not possible to solve this numerical problem with a finer mesh (Fig.~\ref{fig:SI-1:2}\textbf{e}). Instead, higher-order polygon vertices must be used as shown in Figs.~\ref{fig:SI-1:2}\textbf{c}-\textbf{d} to achieve a correct evaluation. In this case, the finite radius of curvature \cite{zhou_minimum_2015} imposed in the topology optimization directly limits the field amplitude at the interface.
We have thus eliminated numerical artefacts from our model as well as in our design but we note that the evaluation of the mode volume at the geometric center of the cavity is in any case a robust quantity \cite{albrechtsen_two_2021}. We show this explicitly by considering the effect of non-vertical sidewalls on the mode volume. Figure~\ref{fig:SI-1:3}\textbf{a} and \textbf{b} compares the normalized electric energy, $\epsilon(r)\abs{\mathbf{E}(r)}^2$, across the bowtie (the $yz$-plane), for vertical and non-vertical sidewalls, respectively. The energy is tightly confined and varies slowly within the dielectric. Note that the energy maximum occurs at the material boundary at $y\neq\SI{0}{nm}$, consistent with the preceding discussions.
Figure~\ref{fig:SI-1:3}\textbf{c} shows the effective mode volume as a function of a global geometry-tuning, where all void features in the $xy$-plane are enlarged or shrunken, which in turn changes the mean bowtie width. Importantly, the mode volume is effectively independent of the sidewall angle when the mode volume is evaluated in the center, $\mathbf{r}_0$, which indicates that this definition is consistent and robust against surface effects.

The robustness of the definition of the mode volume evaluated at the cavity center, which we use in our work, is in contrast to evaluating the mode volume at the maximum of the field or in any other point in space. Figure~\ref{fig:SI-1:3}\textbf{d} and \textbf{e} show $V$ when evaluated at different out-of-plane ($z$) positions at $x=y=0$ for vertical and non-vertical sidewalls, respectively. For a vertical sidewall, the smallest $V$ is at $z=0$. However, when the symmetry is broken by introducing a small deviation from verticality, the mode volume changes and smaller mode volumes seem to occur \cite{saynatjoki_enhanced_2010}. In addition, the mode volume would change significantly when moving the point of evaluation along $y$ towards the material interfaces. This exemplifies some of the inconsistencies that emerge if an arbitrary point, even if it is the point of maximum field intensity, is used to evaluate the mode volume.
The challenges and inconsistencies arising from evaluating the mode volume at the field maximum are particularly important for DBCs because they explicitly rely on field discontinuities at material boundaries but they can also emerge in conventional cavities \cite{almeida_guiding_2004,robinson_ultrasmall_2005,gondarenko_spontaneous_2006,gondarenko_low_2008,schneider_strong_2016,hu_design_2016,choi_self-similar_2017,wang_maximizing_2018,mignuzzi_nanoscale_2019,zhao_minimum_2020,albrechtsen_two_2021}.

To summarize these discussions, the mode volume is a well-defined, robust, and consistent quantity when using Eq.~(1) of the main text and evaluating the mode volume of a quasi-normal mode at the center of the cavity as detailed in refs. \cite{kristensen_generalized_2012,kristensen_normalization_2015}. We stress that this conclusion does not question the validity of the vast majority of previous works on conventional cavities because the field maxima typically occur in the center of such cavities so that the maximum evaluation leads to evaluation at the center. This conclusion does also not rule out the existence of surface effects \cite{sekoguchi_photonic_2014}, which can be very significant albeit hard to control experimentally and potentially governed by divergent fields that cannot be calculated numerically. It would then be crucial to evaluate the mode volume or the LDOS at the frequency, position, and polarization of the emitter for the particular experiment.

\clearpage

\section{Determining the minimum radii of curvature of the nanofabrication}

\begin{figure}[hb!]
    \centering
    \includegraphics[width=.88\linewidth]{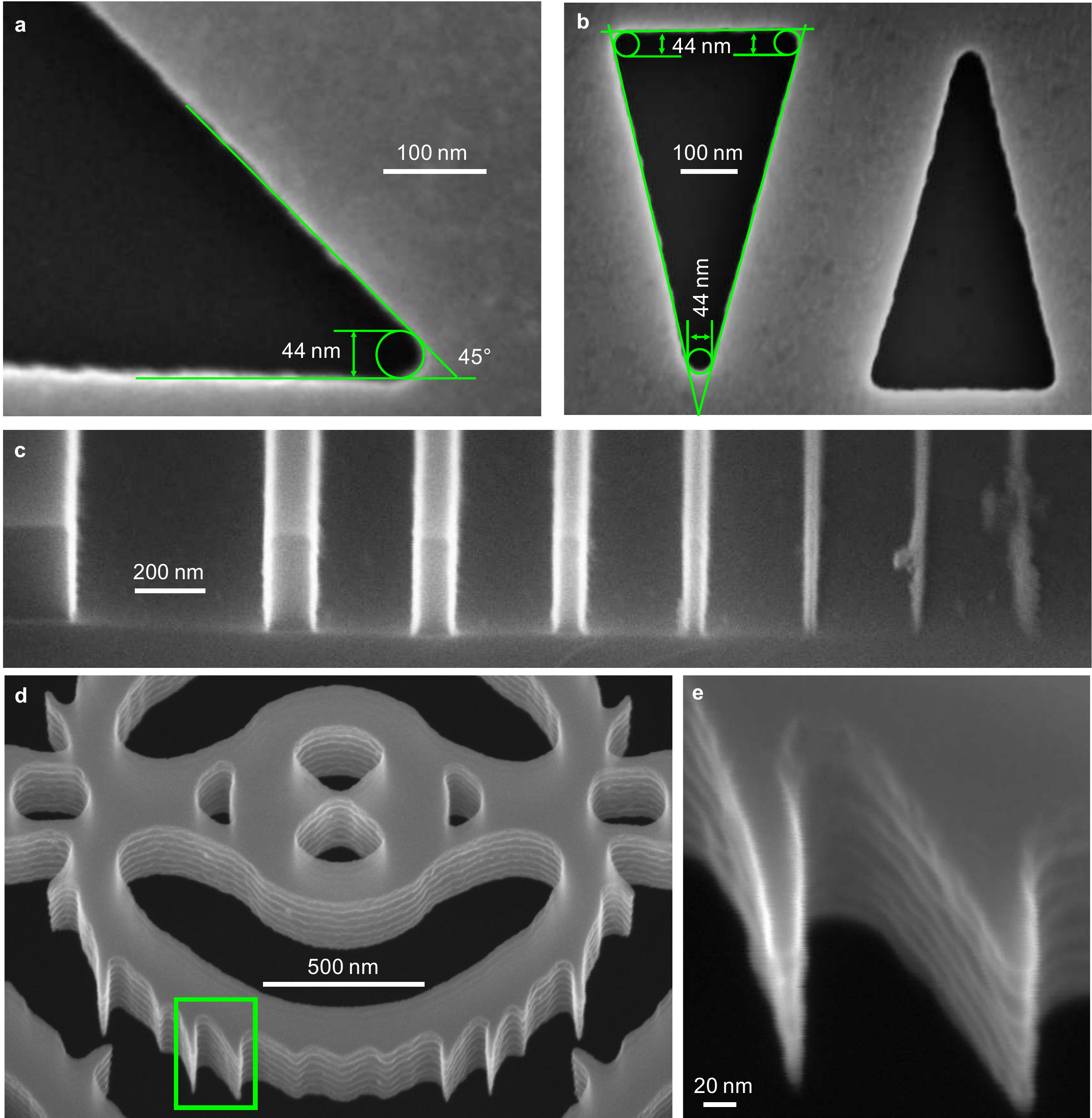}
    \caption{
    \textbf{Measurement of the radii of curvature (ROC) using scanning electron microscopy (SEM) images.}
    \textbf{a-b}, ROC for void features measured to $r_v=\SI{22}{nm}$ on multiple vertices on several triangles of varying size and angles.
    \textbf{c}, Cross-section of \SI{500}{\micro\meter}-long lines of varying width etched in silicon.
    \textbf{d}, Tilted SEM image of a fabricated DBC ($\delta=\SI{-6}{nm}$) where the sharp features in the void limited to $r_s\ge\SI{10}{nm}$ away from the bowtie are visible. The green rectangle highlights the region of the inset shown in \textbf{e}.
    }
    \label{fig:SI-2:1}
\end{figure}

The radius of curvature (ROC) for void features, $r_v$, is evaluated with scanning electron microscopy (SEM) images on corners of fabricated triangles of different sizes as shown in Figs.~\ref{fig:SI-2:1}\textbf{a} and \textbf{b}. In all cases, the ROC for void features is $r_v=\SI{22}{nm}$. The ROC is to a good approximation independent of size and angle but it is easier to measure for larger features.
Figure~\ref{fig:SI-2:1}\textbf{c} shows a cross-section of millimeter-long lines etched with a clear-oxidize-remove-etch dry-etching sequence \cite{nguyen_core_2020,nguyen_ultrahigh_2020,nguyen_cr_2021}, demonstrating the high resolution of the fabrication process. The smallest resolved lines are bent due to electron exposure in the SEM.
Figure~\ref{fig:SI-2:1}\textbf{d} and \textbf{e} shows a DBC with the inset demonstrating that features of $r_s\ge\SI{10}{nm}$ can be realized away from the central bowtie. The bowtie is limited to $r_c\ge\SI{4}{nm}$, which we achieve by carefully modifying the exposure mask locally.

\clearpage

\section{Tuning of exposure mask by global geometry-tuning and local mask corrections}

\begin{figure}[hb!]
    \centering
    \includegraphics[width=\linewidth]{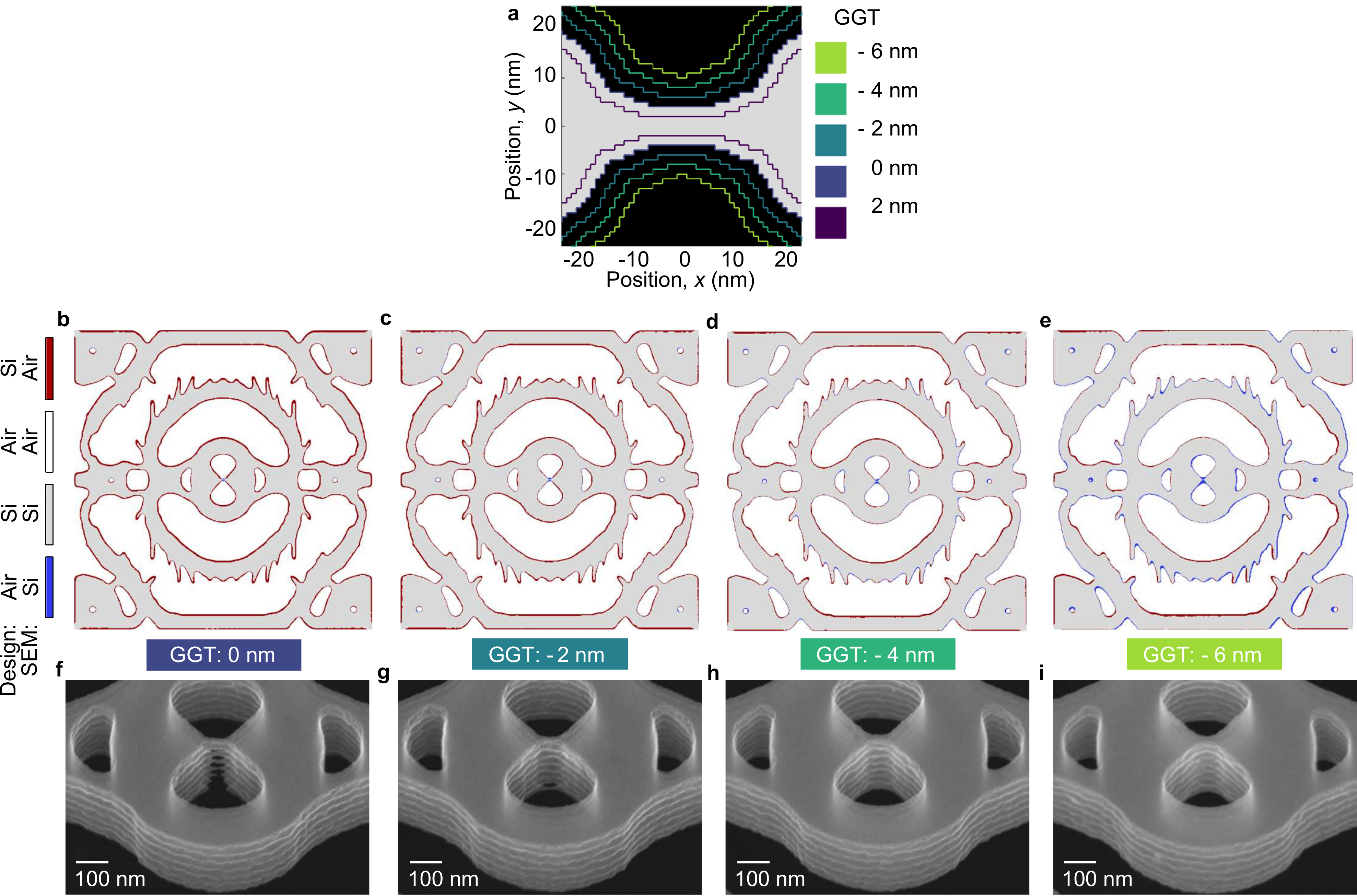}
    \caption{
    \textbf{Evaluation of fabricated DBC from image analysis of scanning electron microscopy (SEM) images.}
    \textbf{a}, Extended global geometry-tuning, $\delta$, from Fig.~1D.
    \textbf{b-e}, Difference between fabricated device and targeted mask obtained by applying a binary filter to a top-view SEM image of a device of each $\delta$ and subtracting the designed mask for each pixel, corresponding to the \SI{40}{\degree} tilted SEM images of the bowtie structures in \textbf{f}-\textbf{i}. \textbf{g-i}, are reproduced from Fig.~1 of the main text for clarity.
    }
    \label{fig:SI-3:1}
\end{figure}

We use a global geometry-tuning to systematically vary the cavity geometry and a local mask correction to fabricate devices with bowtie widths down to the limit set by the critical radius of curvature. To systematically characterize the effects of geometry-tuning and corroborate the observed spectral shifts for different geometry-tunings (Fig.~2 in the main text), we calculate the difference between the original designed mask and binarized top-view scanning electron microscopy (SEM) images of devices with different geometry-tuning, $\delta$, as shown in Figs.~\ref{fig:SI-3:1}\textbf{b}-\textbf{e}.
Figure~\ref{fig:SI-3:1}\textbf{a} shows the change of the exposure mask, where black pixels inside a given contour is exposed with electron-beam lithography. Figures~\ref{fig:SI-3:1}\textbf{f}-\textbf{i} show tilted SEM images of the bowties.
From Figs.~\ref{fig:SI-3:1}\textbf{b}-\textbf{e} it can be seen that the outlines becomes less red and more blue with decreasing $\delta$, indicating that the silicon features grow and the air features shrink as expected. We stress that while the top-view SEM images indicate that the bowties are successfully fabricated, the tilted SEMs in Figs.~\ref{fig:SI-3:1}\textbf{f}-\textbf{i} clearly show that the bowtie in the untuned structure, $\delta=\SI{0}{nm}$, is not resolved. This means that both tilted and top-view images must be compared to evaluate the fabricated bowties, which is challenging when approaching the resolution limit of SEM.

\begin{figure}
    \centering
    \includegraphics[width=.95\linewidth]{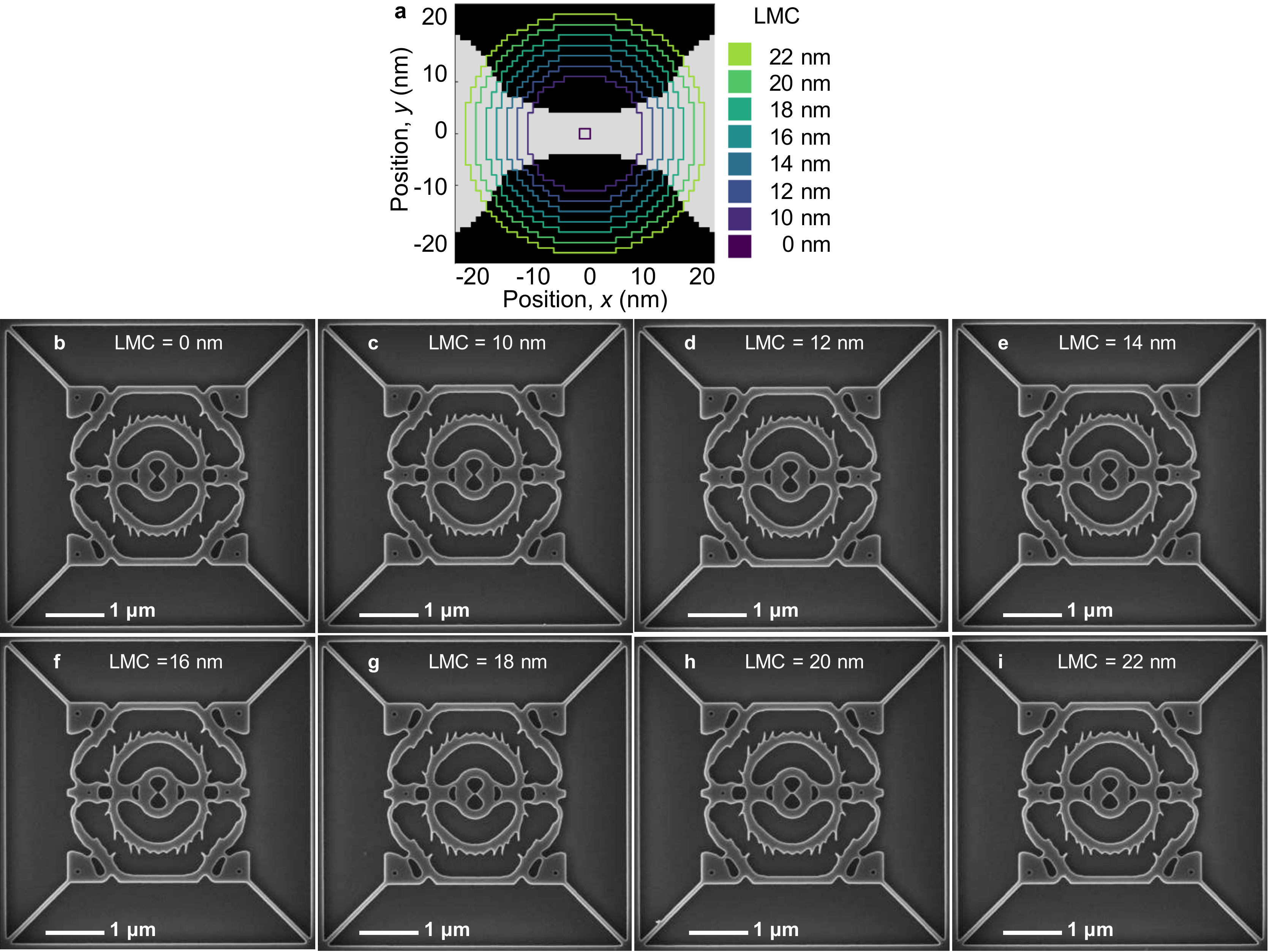}
    \caption{
    \textbf{Local mask corrections (LMC) around the bowtie.}
    \textbf{a}, The outline of the structure (grey) contains features below the radius of curvature but the resolution of critical features can be enhanced by locally removing circular regions from the mask in steps of $\SI{1}{nm\squared}$.
    \textbf{b-i}, Top-view scanning electron micrographs of cavities with no global geometry-tuning, $\delta=\SI{0}{nm}$. This work only considers $\text{LMC}=\SI{22}{nm}$.
    }
    \label{fig:SI-3:5}
\end{figure}

Figure~\ref{fig:SI-3:5}\textbf{a} shows the local mask corrections (LMC) applied to the lithographic exposure mask to achieve the narrow bowtie structure below the process resolution, ultimately limited by the critical radius of curvature $r_c\ge\SI{4}{nm}$.
Figure~\ref{fig:SI-3:5}\textbf{b}-\textbf{i} show top-view SEM images of the 8 local mask corrections before underetching of the structures. In this work we only analyze the device with $\text{LMC}=\SI{22}{nm}$, since this yielded a fully etched bowtie for multiple global geometry-tunings. The global geometry-tuning was applied after the local mask correction.

\clearpage

\section{Quantifying the dimensions of fabricated devices}

\begin{figure}
    \centering
    \includegraphics[width=.95\linewidth]{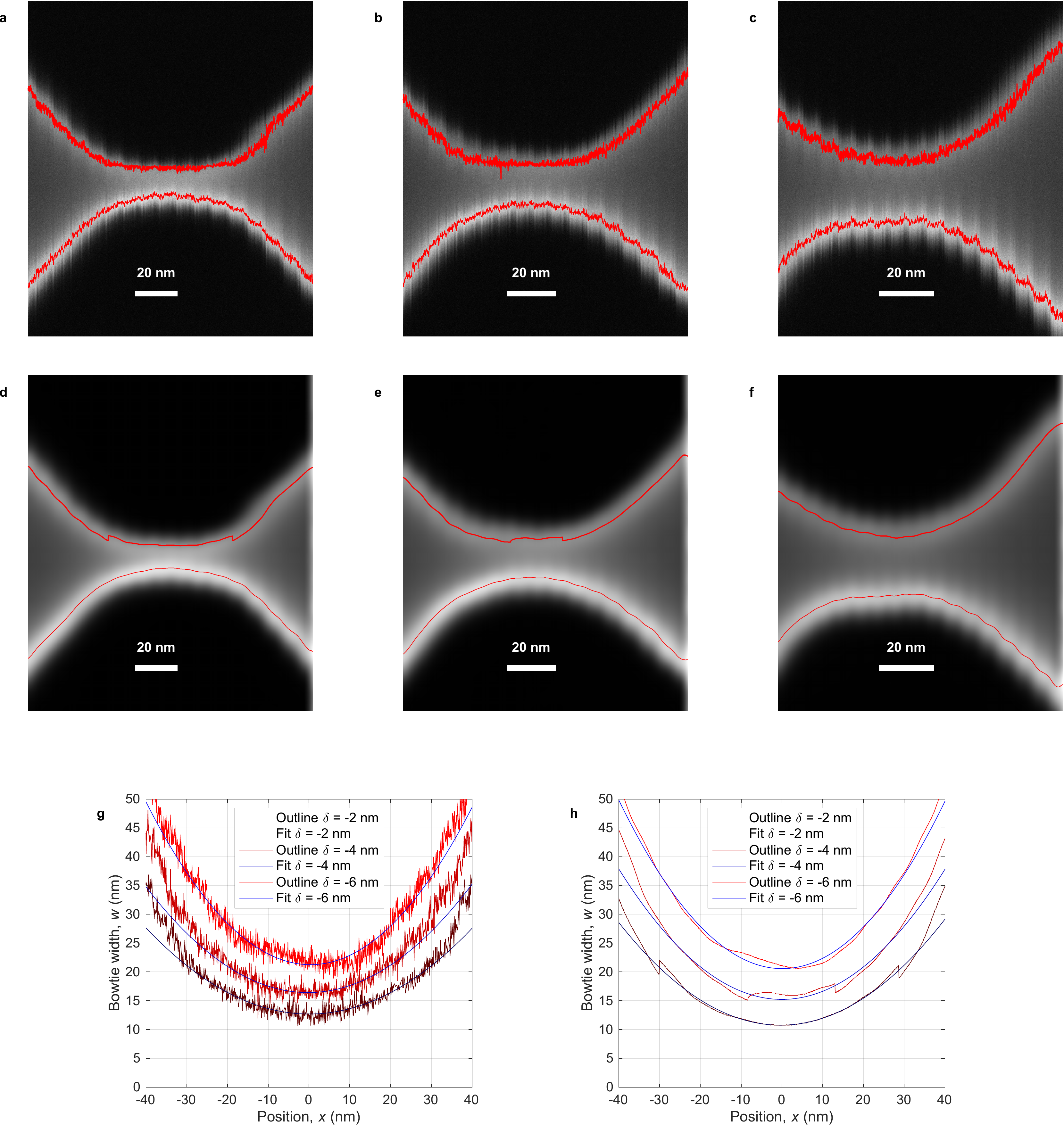}
    \caption{
    \textbf{Top-view scanning electron microscopy (SEM) images of dielectric bowties.}
    \textbf{a-c}, Top-view SEM images acquired with a secondary electron detector under \SI{25}{keV} excitation on nominally identical device copy~5 (with optical spectra shown in Fig.~2 in the main text) for global geometry-tuning, $\delta=\SI{-2}{nm}$, $\delta=\SI{-4}{nm}$, and $\delta=\SI{-6}{nm}$, respectively. The red lines traces the peak intensity in the SEM images and the distance between them measures the width.
    \textbf{d-f}, Smoothed SEMs from \textbf{a-c} using a Gaussian filter with 25 pixels variance.
    \textbf{g-h}, Widths obtained from \textbf{a-c} and \textbf{d-f}, respectively. The dark-red lines shows the measurements and the dark-blue lines a corresponding quadratic fit around the central part, yielding bowtie widths of \SI{13}{nm}, \SI{15}{nm}, and \SI{21}{nm}, respectively for \textbf{d-f}, which combined with the small sidewall angle results in average widths (at the center of the membrane) of $(8\pm5)\ \SI{}{nm}, (10\pm5)\ \SI{}{nm}, \text{and}\ (16\pm5)\ \SI{}{nm}$, respectively.
    }
    \label{fig:SI-3:3}
\end{figure}

\begin{figure}
    \centering
    \includegraphics[width=.8\linewidth]{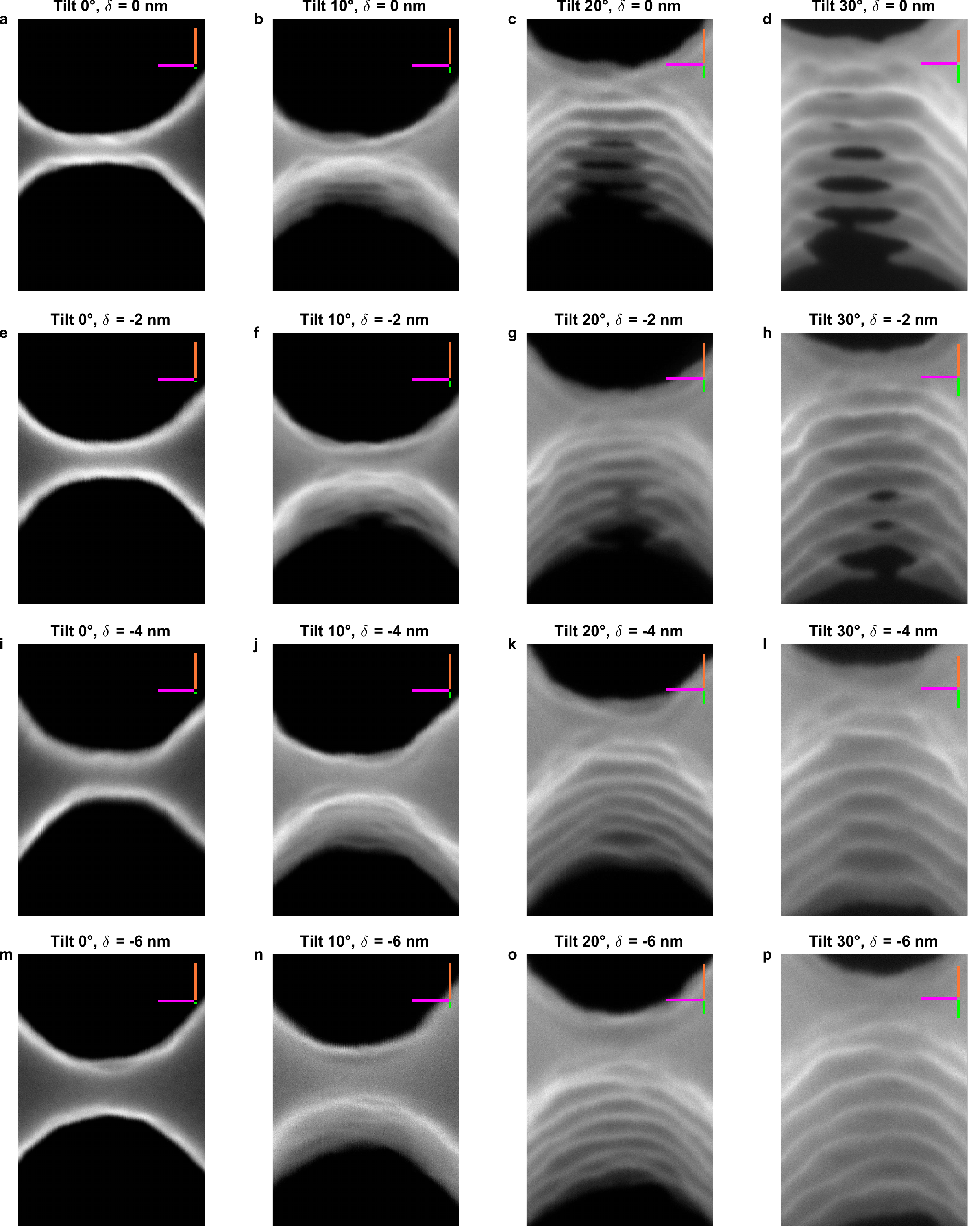}
    \caption{
    \textbf{Array of scanning electron microscopy (SEM) images of geometry-tuned cavities for different viewing angles.}
    \textbf{a-d}, Shows \SI{0}{\degree}, \SI{10}{\degree}, \SI{20}{\degree}, and \SI{30}{\degree} tilted SEM images for global geometry-tuning $\delta=\SI{0}{nm}$, where the bowtie is resolved but no resonance is measured in the far-field measurements.
    \textbf{e-h}, \textbf{i-l}, and \textbf{m-p}, Tilted images similar to \textbf{a-d} but for $\delta=\SI{-2}{nm}$, $\delta=\SI{-4}{nm}$, and $\delta=\SI{-6}{nm}$, respectively, showing devices where the optical mode is observed in far-field measurements (see Fig.~2 in the main text).
    All images are measured using a \SI{19}{keV} electron beam with an in-lens detector collecting secondary electrons on nominally identical device copy~6.
    All scale bars are \SI{20}{nm}, with the magenta, orange, and green scale bars showing the length along the $x$, $y$, and $z$ directions, respectively. The green scale bars in \textbf{l} and \textbf{p} enables verification that the device layer is \SI{240}{nm} thick, etched with 8 scallops that are each $\sim\SI{30}{nm}$ tall.
    }
    \label{fig:SI-3:4}
\end{figure}

Figure~\ref{fig:SI-3:3}\textbf{a-c} shows top-view scanning electron microscopy (SEM) images with $<\SI{100}{pm}$ pixels where the scan-direction is perpendicular to the bowtie (along the $y$-direction) to minimize drift errors in the measurement. The discontinuous outlines results as the dimensions are small compared to the resolution of the electron microscope.
Figure~\ref{fig:SI-3:3}\textbf{d-f} shows smoothed images to achieve a smooth outline of the top-view of the fabricated bowtie.
We compute the width for each vertical line individually as shown in Figure~\ref{fig:SI-3:3}\textbf{g} and \textbf{h} and fit a quadratic function around the center. A quadratic function is chosen since the topology optimization was tolerance-constrained to a void radius of curvature $r_v=\SI{22}{nm}$.
Although there are several nanometers of fluctuations in the widths in Fig.~\ref{fig:SI-3:3}\textbf{g}, it is substantially lower than the fluctuations on either side in Fig.~\ref{fig:SI-3:3}\textbf{a-c}, indicating that the scan fluctuations are correlated.

Figure~\ref{fig:SI-3:4} shows an array of different tilted views of different geometry-tuned devices from the same nominally identical copy (see optical spectra in Fig.~2 in the main text). This reveals that top-view SEM images alone are insufficient to characterize our cavities, as they do not reveal if the structure is correctly etched. We note that these measurements are consistent with the value of the device-layer thickness, $t=\SI{240}{nm}$, which we measured with high precision using variable-angle ellipsometric spectroscopy before the topology optimization.
These images also enable us to identify that the bowtie is etched during the first 8 cycles of the cyclic dry-etching process with each cycle etching $\sim\SI{30}{nm}$, while 10 cycles are needed to etch the small air holes fully due to etch lag \cite{jansen_bsm_1997}.
From Fig.~\ref{fig:SI-3:4}\textbf{l} and \textbf{p} it can be seen that the process is free of notching \cite{hwang_origin_1997}. From Fig.~\ref{fig:SI-3:4}\textbf{g-h} and \textbf{k}, we estimate that the thickness of the bowtie is more narrow at the bottom, approximately \SI{10}{nm} narrower, which indicates a sidewall negativity $\sim\SI{1}{\degree}\pm\SI{0.5}{\degree}$.
Comparing Fig.~\ref{fig:SI-3:4}\textbf{g-h} of nominally identical copy~5 to the tilted image in Fig.~1\textbf{e} in the main text of nominally identical copy~3 we note that the scallops penetrates the bowtie more places in copy~5 compared to copy~3. However, from the far-field measurements in Fig.~2 in the main text we observe a clear resonance for both cavities with resonant wavelength $\lambda_0\approx\SI{1440}{nm}$.
Figure~\ref{fig:SI-3:4}\textbf{h} shows that the scallops just penetrates the bowtie at the 6th scallop indicating a scallop depth of $\sim\SI{3}{nm}$.
Lastly, the \SI{20}{nm} scale bars along $y$, and $z$ are obtained by scaling the scale bar along $x$ by $\cos(\theta)$ and $\sin(\theta)$, respectively, with $\theta$ the tilt angle.

\clearpage

\section{Scattering scanning near-field optical microscopy measurements}

\begin{figure}[b!]
    \centering
    \includegraphics[width=.8\textwidth]{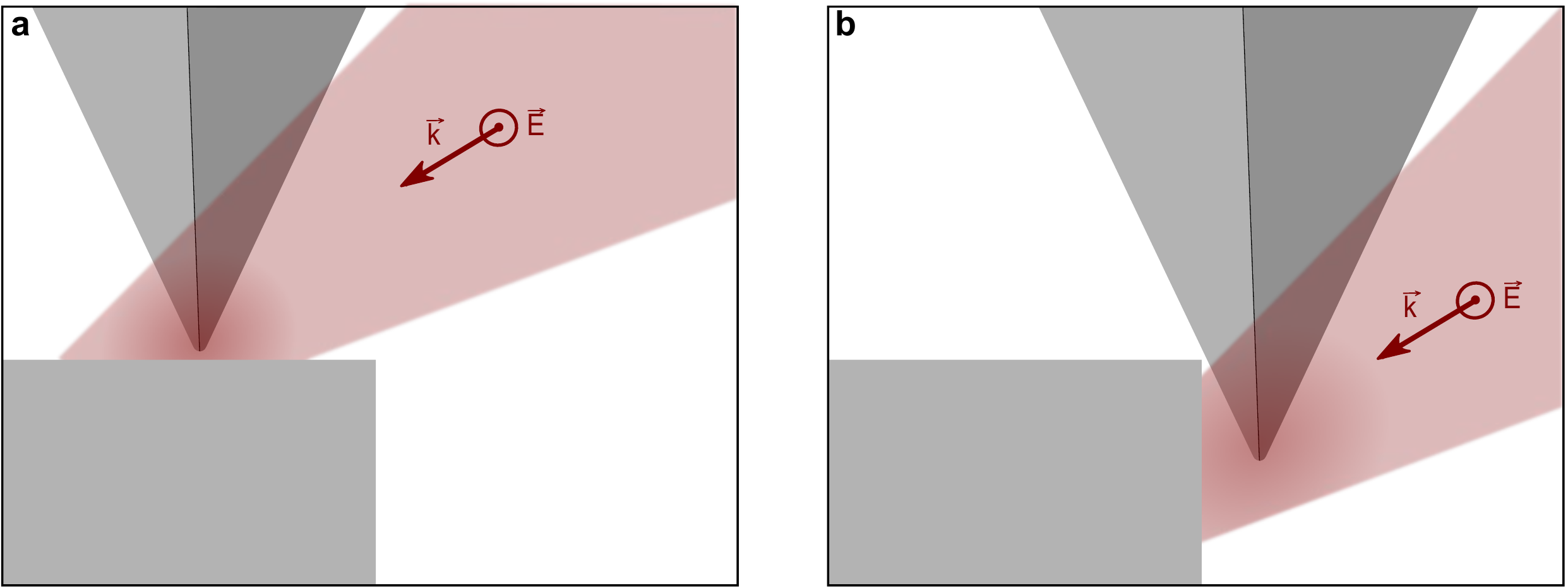}
    \caption{
    \textbf{Schematic of atomic force microscope (AFM) tip under optical illumination for scattering near-field optical microscopy}.
    \textbf{a}, The AFM tip is made of silicon and has a nominal radius of $r\sim\SI{10}{nm}$. It is operated in tapping mode and data is recorded when the tip is $\sim\SI{5}{nm}$ above the surface. The laser is s-polarized and has a diffraction-limited spot size of approximately \SI{2}{\micro\meter} as indicated by the red fields and vectors. \textbf{b}, The measurements are governed by interactions between the sidewalls of the tip and the nanostructure when the tip is above void features of the geometry and we therefore discard these regions in our analysis of the measured mode.
    }
    \label{fig:sSNOM-schematic}
\end{figure}

We perform near-field measurements to experimentally verify the optical field confinement to a single hotspot in the middle of the cavities. To show this explicitly, this section presents near-field measurements on devices with different geometry-tunings, which results in different resonance wavelengths but essentially identical near-field maps. Figure~\ref{fig:sSNOM-schematic} shows a schematic of the atomic force microscope (AFM) tip used for the near-field measurement. This gives reliable and detailed information about the near-field when scanning above the sample. However, when moving into a void feature, complex perturbations of the scattered light emerge and we discard data obtained in this regime for our final analysis. The excitation laser is s-polarized to minimize the excitation of the tip \cite{novotny_principles_2012} and s-polarization of the scattered near-field is detected using a polarizer, as has been done in previous works \cite{schnell_phase-resolved_2010,kim_polarization-selective_2008}. 

\begin{figure}
    \centering
    \includegraphics[width=\textwidth]{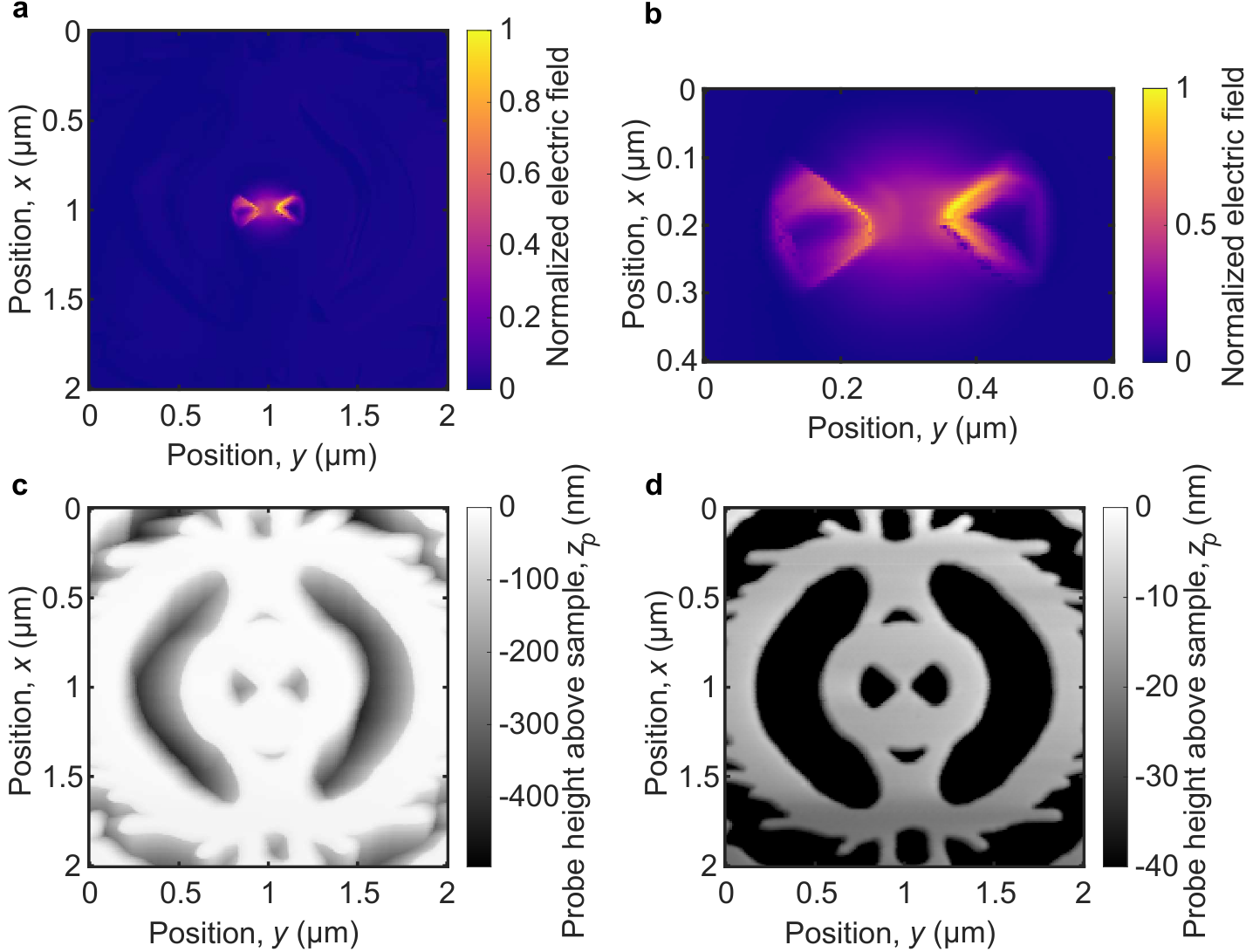}
    \caption{
    \textbf{Near-field mapping of a cavity with global geometry-tuning $\bm{\delta=}\SI{-4}{nm}$}. \textbf{a}-\textbf{b}, Near-field amplitude, $\abs{\bm{E}}$, of the cavity excited on-resonance, $\lambda_0 = \SI{1489.4}{nm}$. \textbf{c-d}, Atomic force microscope (AFM) signal of the cavity, obtained from the same measurement, with \textbf{c} showing the full range where the tip moves through the air holes, and \textbf{d}, truncated at $z_p=-\SI{40}{nm}$ to highlight the surface.
    }
    \label{fig:Maps1}
\end{figure}

\begin{figure}
    \centering
    \includegraphics[width=\textwidth]{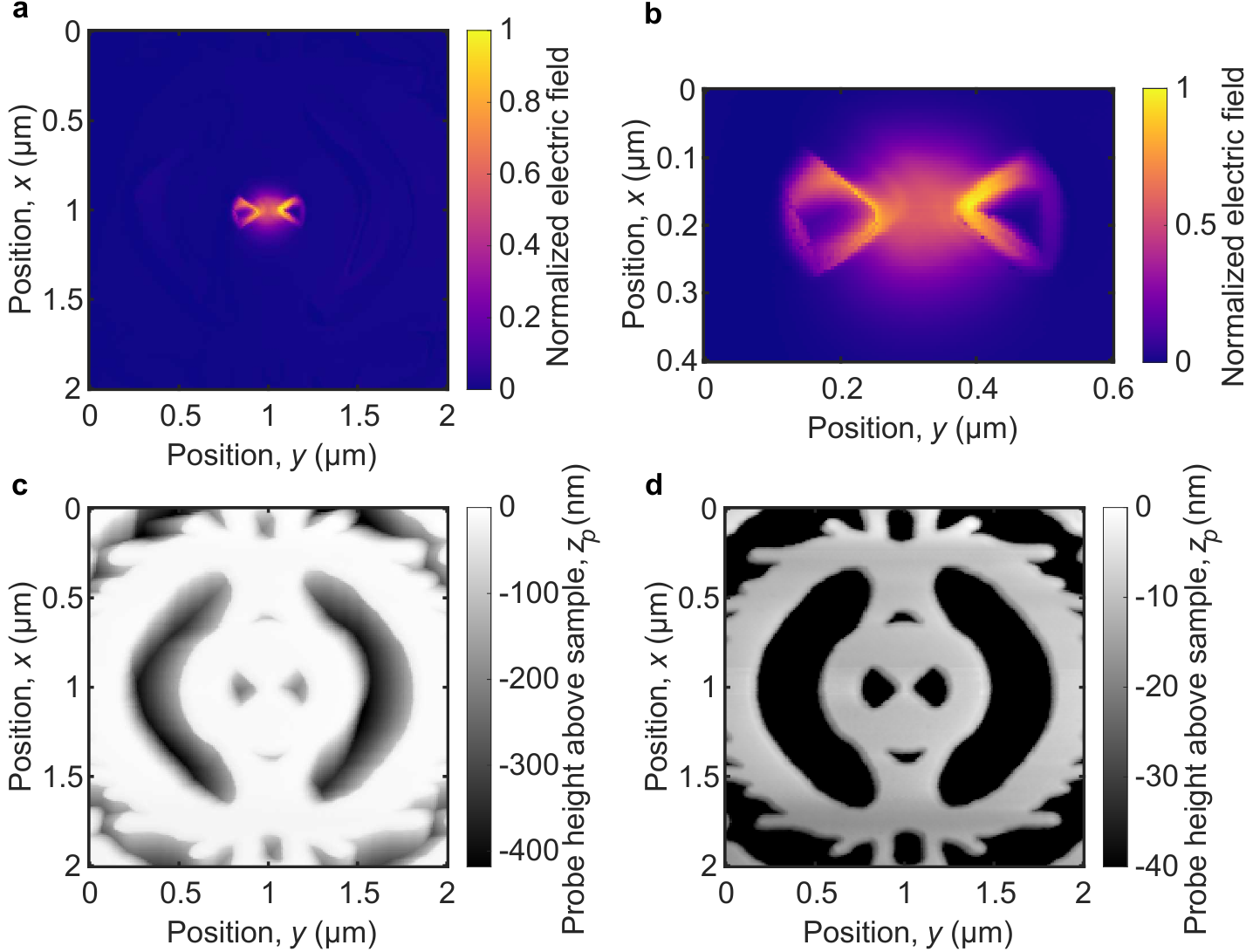}
    \caption{
    \textbf{Near-field mapping of a cavity with global geometry-tuning $\bm{\delta=}\SI{-6}{nm}$}. \textbf{a}-\textbf{b}, Near-field amplitude, $\abs{\bm{E}}$, of the cavity excited on-resonance, $\lambda_0 = \SI{1522.9}{nm}$. \textbf{c-d}, Atomic force microscope (AFM) signal of the cavity, obtained from the same measurement, with \textbf{c} showing the full range where the tip moves through the air-holes, and \textbf{d}, truncated at $z_p=-\SI{40}{nm}$ to highlight the surface.
    }
    \label{fig:Maps2}
\end{figure}

Figures~\ref{fig:Maps1}\textbf{a} and \textbf{b} show two measurements of the near-field amplitude for a cavity with global geometry-tuning $\delta=-\SI{4}{nm}$ recorded on resonance, $\lambda_0=\SI{1489.4}{nm}$, with two different spatial resolutions. This demonstrates that the field is confined in the center with excellent suppression of the background yielding strong correlation with the simulation as discussed in the main text. The contours of high field strengths shaped like the cavity holes on both sides of the center stem from the fact that the tip is moving inside the void features, which is consistent with the AFM measurement shown in Fig.~\ref{fig:Maps1}\textbf{c}.
The sharp silicon edges appear sloped as the tip shaft prevents the tip from going further down inside a hole. For clarity, Fig.~\ref{fig:Maps1}\textbf{d} shows the same AFM map truncated to show only a thickness range of \SI{40}{nm}.
The same measurements and data analysis have been performed for the cavity with global geometry-tuning, $\delta=-\SI{6}{nm}$, shown in Figure~\ref{fig:Maps2}, which yield similar results.

\begin{figure}
    \centering
    \includegraphics[width=\textwidth]{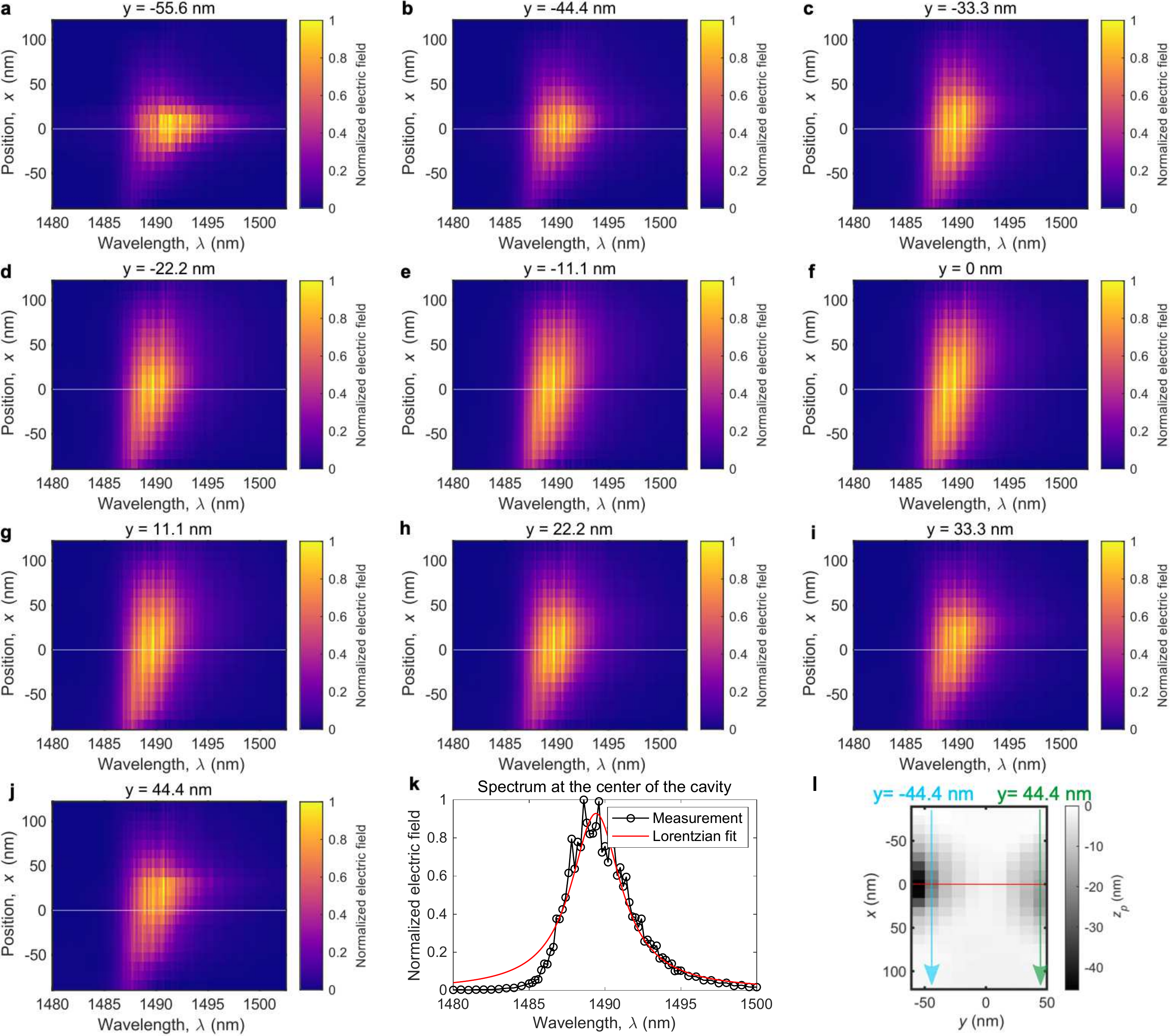}
    \caption{
    \textbf{Near-field spectrum of a cavity with global geometry-tuning $\bm{\delta=}\SI{-4}{nm}$}.
    \textbf{a}-\textbf{j}, s-SNOM spectra at different wavelengths for fixed $y$ positions, scanning along the bowtie. The white line indicates the $x$ position of the cavity center.
    \textbf{k}, Spectrum at the cavity center with a Lorentzian fit.
    \textbf{l}, Example of an AFM map obtained during the s-SNOM measurements. The red line indicates the $x$ position of the cavity center.
    }
    \label{fig:Spec1}
\end{figure}

\begin{figure}
    \centering
    \includegraphics[width=\textwidth]{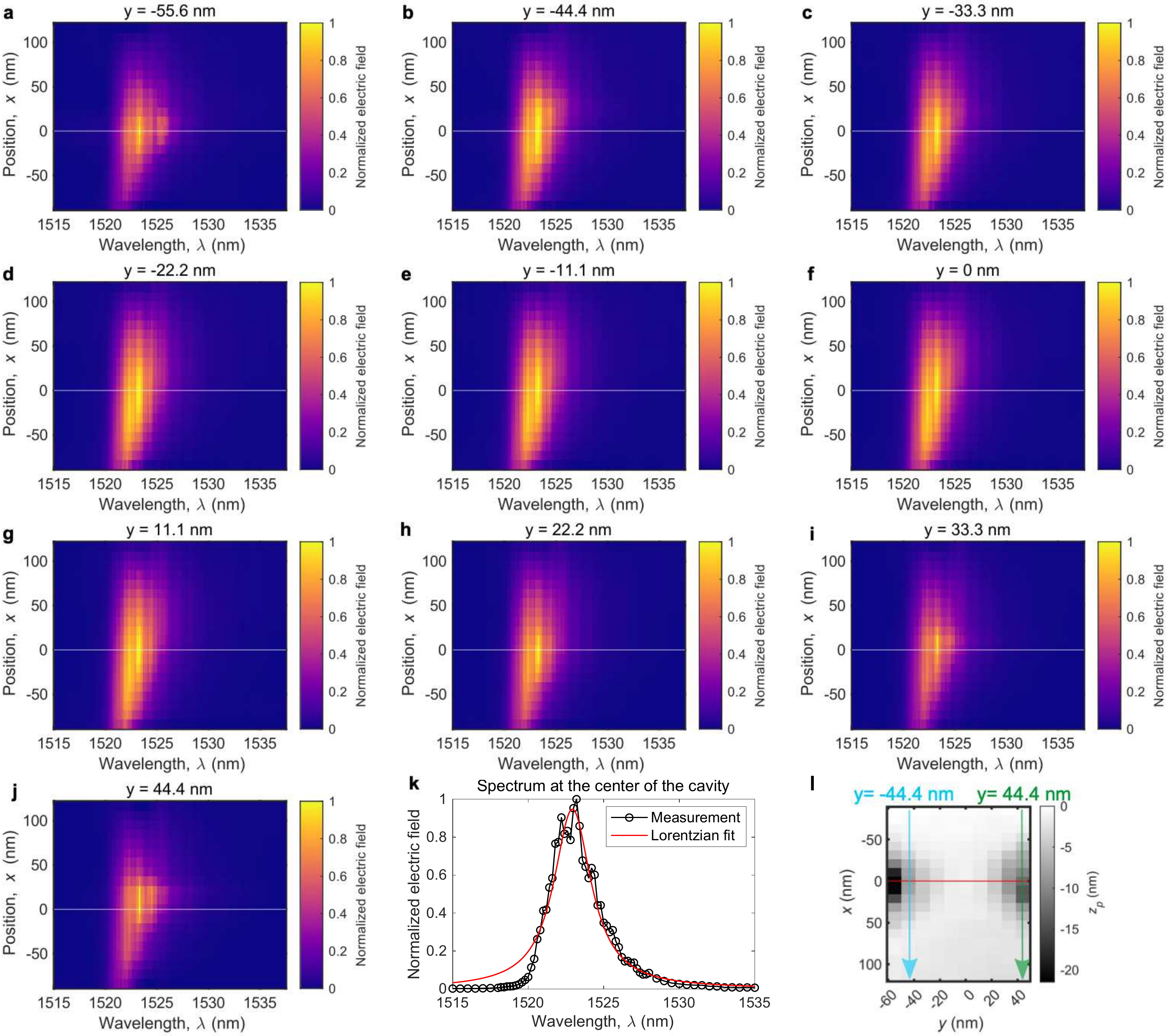}
    \caption{
    \textbf{Near-field spectrum of a cavity with global geometry-tuning $\bm{\delta=}\SI{-6}{nm}$}.
    \textbf{a}-\textbf{j}, s-SNOM spectra at different wavelengths for fixed $y$ positions, scanning along the bowtie. The white line indicates the $x$ position of the cavity center.
    \textbf{k}, Spectrum at the cavity center with Lorentzian fit.
    \textbf{l}, Example of an AFM map taken for the spectra measurements. The red line indicates the $x$ position of the cavity center.
    }
    \label{fig:Spec2}
\end{figure}

To map the resonance of the cavities in the near-field, we sweep the wavelength of the tunable excitation laser as discussed in Methods. A small map of the center of the cavity was acquired for wavelength steps of \SI{0.5}{nm}, with an extra sampling spaced \SI{0.2}{nm} around the resonance peak, in a range of \SI{20}{nm} around the expected resonance from the far-field measurements. The results from these measurements are summarized in Fig.~\ref{fig:Spec1} and \ref{fig:Spec2}. 
Figures \ref{fig:Spec1}\textbf{a}-\textbf{j} present the spectra for the cavity with global geometry-tuning, $\delta=-\SI{4}{nm}$, along a line of varying $x$ position and fixed $y$ position. Most of the spectra have a resonance peak around \SI{1489.4}{nm}.

The spectrum in the middle of the cavity is shown in Fig.~\ref{fig:Spec1}\textbf{k}. We perform a fit with a Lorentzian in the frequency domain,
\begin{equation}
    L(\omega) \propto \frac{\Gamma/2}{(\omega-\omega_0)^2+(\Gamma/2)^2},
\end{equation}
which gives a near-field quality factor, $Q_\text{NF}=\omega_0/\Gamma = 370 \pm 40$ around the resonant wavelength $\lambda_0 = (1489.4 \pm 0.1)\ \SI{}{nm}$. The near-field quality factor is lower than the quality factor measured in the far-field, which we attribute to losses induced by the presence of the tip \cite{gotzinger_towards_2001}.

Figure~\ref{fig:Spec1}\textbf{l} shows an example of an AFM map acquired during the frequency sweep, here at the wavelength of $\lambda_0 = \SI{1480}{nm}$. The blue arrow indicates the pixels selected to plot the spectrum in Fig.~\ref{fig:Spec1}\textbf{b} and the green arrow shows the pixels for Fig.~\ref{fig:Spec1}\textbf{j}. The red line indicates the $x$ position of the cavity center.
The same measurements and data treatment have been performed for the cavity of global geometry-tuning, $\delta=-\SI{6}{nm}$. These are shown in Fig.~\ref{fig:Spec2}. We extract $\lambda_0 = (1522.9 \pm 0.1)\ \SI{}{nm}$ from a Lorentzian fit and calculate the near-field quality factor $Q_\text{NF} = 510 \pm 50$.

\begin{figure}
    \centering
    \includegraphics[width=\linewidth]{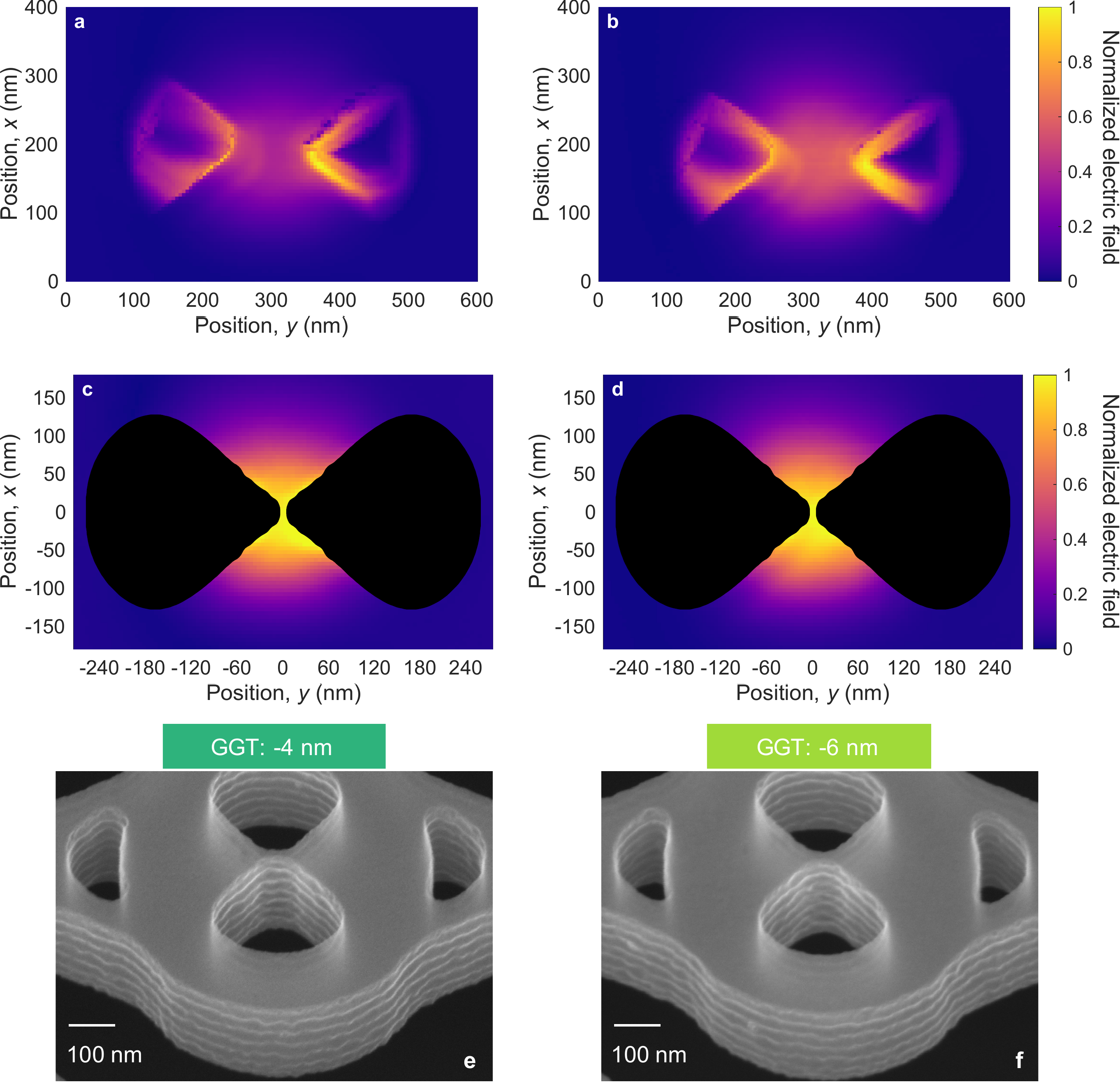}
    \caption{
    \textbf{Comparison of near-field measurements of cavities with different global geometry-tuning, $\bm{\delta}$.}
    \textbf{a}-\textbf{b}, Normalized electric field amplitude obtained by s-SNOM measurements of DBCs with $\delta=\SI{-4}{nm}$ and $\delta=\SI{-6}{nm}$, respectively. Each map is normalized to the maximum measured.
    \textbf{c}-\textbf{d}, Same measurements as \textbf{a}-\textbf{b} but centered and with void features blacked out and normalized to the value at the center, emphasizing that the measurements are near identical for different bowtie widths.
    \textbf{e}-\textbf{f}, Scanning electron micrographs of the cavity bowties measured with s-SNOM showing the difference in bowtie widths as discussed in the main text.
    }
    \label{fig:SI-4:1}
\end{figure}

Figure~\ref{fig:SI-4:1} shows a comparison between the dielectric bowtie cavities of different global geometry-tunings. Figures~\ref{fig:SI-4:1}\textbf{a}-\textbf{b} are normalized to the maximum intensity in the maps, and Figs.~\ref{fig:SI-4:1}\textbf{c}-\textbf{d} are normalized to the center with the air-regions blacked out to qualitatively assess the similarity between the measurements irregardless of the different bowtie widths as indicated by the tilted scanning electron microscopy (SEM) images in Figs.~\ref{fig:SI-4:1}\textbf{e}-\textbf{f}. The location of the blacked-out region (i.e. the center of the map) is determined as the position yielding the best overlap between theory and experiment.
The Bhattacharyya coefficient \cite{bhattacharyya_measure_1943} for the overlap between the two maps in Figs.~\ref{fig:SI-4:1}\textbf{a}-\textbf{b} is $t=0.985$, and for the two blacked out maps in Figs.~\ref{fig:SI-4:1}\textbf{c}-\textbf{d} it is $t=0.996$. The overlap between the measurements for $\delta=\SI{-4}{nm}$ shown in Fig.~\ref{fig:Maps1}\textbf{b} and the calculated mode is $t=0.984$ when the calculated mode is convolved with a Gaussian instrument function with $\sigma=(37\pm5)\ \SI{}{nm}$, consistent with the results of the main text.

\clearpage

\section{Overview of experimental data in this work}
The present paper reports experimental results of a sample with 336 dielectric bowtie cavities, split across six nominally identical copies of 56 different cavities. Each of the 56 cavities are different due to systematic modifications of the exposure mask according to the following procedure: First, the mask is modified by a local mask correction (LMC) around the central bowtie, see Fig.~\ref{fig:SI-3:5}, and second, the mask is tuned by a global geometry-tuning ($\delta$), see Fig.~\ref{fig:SI-3:1}.
All figures, with the exception of Fig.~\ref{fig:SI-3:5}, report data of cavities with $\text{LMC}=\SI{22}{nm}$. Table~\ref{tab:dataOverview} lists all figures that present experimental data as well as the corresponding device (copy and global geometry-tuning) used to produce it.
The near-field optical measurements are not performed on devices with $\delta=\SI{-2}{nm}$, corresponding to a mean bowtie bridge width, $w=(8\pm5)~\SI{}{nm}$, since their resonance at $\lambda_0\sim\SI{1440}{nm}$ lies outside the range of the s-SNOM excitation laser, see Methods.

\begin{table}[bh!]
    \centering
    \begin{tabular}{|l|c|}
        \hline
        Figure & Device description \\
        \hline
        Fig. 1c & Copy 3, $\delta=\SI{-2}{nm}$. \\
        Fig. 1e & Copy 3, $\delta=\SI{-2}{nm}$. \\
        Fig. 1f & Copy 3, $\delta=\SI{-4}{nm}$. \\
        Fig. 1g & Copy 3, $\delta=\SI{-6}{nm}$. \\
        \hline
        Fig. 2a & Copy 5, $\delta=\SI{-6}{nm}$. \\
        Fig. 2b & Copy \{1-6\}, $\delta=\SI{-2}{nm}$. \\
        Fig. 2c & Copy \{1-6\}, $\delta=\SI{-4}{nm}$. \\
        Fig. 2d & Copy \{1-6\}, $\delta=\SI{-6}{nm}$. \\
        Fig. 2e & Mean and standard deviation across all 18 cavities. \\
        \hline
        Fig. 3a to d & Copy 3, $\delta=\SI{-4}{nm}$. \\
        \hline
        Fig. S5\textbf{a}-\textbf{b} & Precursor sample 1 to quantify the radius of curvature.\\
        Fig. S5\textbf{c} & Precursor sample 2 to quantify dry-etch performance. \\
        Fig. S5\textbf{d}-\textbf{e} & Copy 3, $\delta=\SI{-6}{nm}$. \\
        \hline
        Fig. S6\textbf{b},\textbf{f} & Copy 3, $\delta=\SI{0}{nm}$. \\
        Fig. S6\textbf{c},\textbf{g} & Copy 3, $\delta=\SI{-2}{nm}$. \\
        Fig. S6\textbf{d},\textbf{h} & Copy 3, $\delta=\SI{-4}{nm}$. \\
        Fig. S6\textbf{e},\textbf{i} & Copy 3, $\delta=\SI{-6}{nm}$. \\
        \hline
        Fig. S7\textbf{b}-\textbf{i} & Copy 1, $\delta=\SI{0}{nm}$, all LMC. \\
        \hline
        Fig. S8\textbf{a},\textbf{d} & Copy 6, $\delta=\SI{-2}{nm}$. \\
        Fig. S8\textbf{b},\textbf{e} & Copy 6, $\delta=\SI{-4}{nm}$. \\
        Fig. S8\textbf{c},\textbf{f} & Copy 6, $\delta=\SI{-6}{nm}$. \\
        Fig. S8\textbf{g}-\textbf{h} & Copy 6, all $\delta$. \\
        \hline
        Fig. S9\textbf{a}-\textbf{d} & Copy 6, $\delta=\SI{0}{nm}$. \\
        Fig. S9\textbf{e}-\textbf{h} & Copy 6, $\delta=\SI{-2}{nm}$. \\
        Fig. S9\textbf{i}-\textbf{l} & Copy 6, $\delta=\SI{-4}{nm}$. \\
        Fig. S9\textbf{m}-\textbf{p} & Copy 6, $\delta=\SI{-6}{nm}$. \\
        \hline
        Fig. S11\textbf{m}-\textbf{p} & Copy 3, $\delta=\SI{-4}{nm}$. \\
        \hline
        Fig. S12\textbf{m}-\textbf{p} & Copy 3, $\delta=\SI{-6}{nm}$. \\
        \hline
        Fig. S13\textbf{m}-\textbf{p} & Copy 3, $\delta=\SI{-4}{nm}$. \\
        \hline
        Fig. S14\textbf{m}-\textbf{p} & Copy 3, $\delta=\SI{-6}{nm}$. \\
        \hline
        Fig. S15\textbf{a},\textbf{c},\textbf{e} & Copy 3, $\delta=\SI{-4}{nm}$. \\
        Fig. S15\textbf{b},\textbf{d},\textbf{f} & Copy 3, $\delta=\SI{-6}{nm}$. \\
        \hline
    \end{tabular}
    \caption{\textbf{Overview of data, figures, and results with corresponding devices.} All data except Fig.~\ref{fig:SI-3:5} reports on devices with local mask correction, $\text{LMC}=\SI{22}{nm}$.}
    \label{tab:dataOverview}
\end{table}

\clearpage

\printbibliography[filter=onlysuppinfo]

\end{document}